\documentclass[11pt,letterpaper]{article}

\usepackage[margin=1in]{geometry}
\usepackage[T1]{fontenc}
\usepackage{lmodern}
\usepackage{microtype}
\usepackage{graphicx}
\usepackage{natbib}
\usepackage{amsmath,amssymb,amsfonts,bm}
\usepackage{xfrac}
\usepackage{nicefrac}
\usepackage{xcolor}
\usepackage{appendix}
\usepackage[normalem]{ulem}
\usepackage{enumitem}
\usepackage{algorithm}
\usepackage{algpseudocode}
\usepackage{anyfontsize}
\usepackage{subcaption}
\usepackage{comment}
\usepackage{booktabs}
\usepackage{tabularx}
\usepackage{array}
\usepackage{multirow}
\usepackage{threeparttable}
\usepackage{makecell}
\usepackage{listings}
\usepackage{tcolorbox}
\tcbuselibrary{listings}
\usepackage[hidelinks]{hyperref}

\newcolumntype{Y}{>{\centering\arraybackslash}X}
\newcolumntype{C}[1]{>{\centering\arraybackslash}m{#1}}

\lstdefinestyle{splat}{
  language=Matlab,
  basicstyle=\ttfamily\footnotesize,
  keywordstyle=\color{black}\bfseries,
  commentstyle=\color{black!60},
  stringstyle=\color{black},
  columns=fullflexible,
  keepspaces=true,
  showstringspaces=false,
  frame=single,
  rulecolor=\color{black!30},
  framerule=0.4pt,
  framesep=5pt,
  aboveskip=5pt,
  belowskip=0pt,
  breaklines=false,
  morekeywords={gaussFilter,delta}
}

\usepackage{tcolorbox}
\tcbuselibrary{listings}

\newtcblisting{splatCode}{
    listing only,
    colback=black!15,
    colframe=black!25,
    boxrule=0.4pt,
    sharp corners,
    boxsep=0pt,
    left=6pt,
    right=6pt,
    top=5pt,
    bottom=5pt,
    before skip=0pt,
    after skip=0pt,
    listing options={
        style=splat,
        basicstyle=\ttfamily\scriptsize,
        backgroundcolor={},
        frame=none,
        aboveskip=0pt,
        belowskip=0pt,
        columns=fullflexible,
        keepspaces=true
    }
}

\newcommand{\splatHeader}[3]{%
\begin{tcolorbox}[
    width=\linewidth,
    colback=#1!10,
    colframe=#1!10,
    boxrule=0pt,
    sharp corners,
    boxsep=0pt,
    left=0pt,
    right=0pt,
    top=2pt,
    bottom=2pt,
    before skip=0pt,
    after skip=0pt
]
\centering
{\footnotesize\bfseries\boldmath
    \textcolor{#1!65!black}{#2:}
    \enspace
    \(\textstyle #3\)\par
}
\end{tcolorbox}%
}

\newcommand{\modelSEG}[1]{\ensuremath{\text{\textsc{SEG}}_{#1\mathrm{mm}}}}
\newcommand{\modelBase}{\textsc{Base }}
\newcommand{\modelDOI}{\textsc{$^+$DOI }}
\newcommand{\modelEnergy}{\textsc{$^+$Energy }}
\newcommand{\modelDepth}{\textsc{$^+$Depth }}

\newcommand{\modelwidth}{1.3cm}
\newcommand{\numberwidth}{1.1cm}
\newcommand{\sourcegap}{0.35cm}

\title{SPLAT: An Event-Specific System Response Model for\\ Continuous-Domain List-Mode TOF PET}
\author{Ronglong Fang\textsuperscript{1}, Andrzej Krol\textsuperscript{2},
Yuesheng Xu\textsuperscript{3}, Eric S. Harmon\textsuperscript{4},\\
Michael O. Thompson\textsuperscript{5}, Wei Zheng\textsuperscript{2},
C. Ross Schmidtlein\textsuperscript{1}\thanks{Corresponding author: schmidtr@mskcc.org}\\[0.7em]
\small \textsuperscript{1}Department of Medical Physics, Memorial Sloan Kettering Cancer Center, New York, NY, USA\\
\small \textsuperscript{2}Departments of Radiology, SUNY Upstate Medical University, Syracuse, NY, USA\\
\small \textsuperscript{3}Departments of Mathematics and Statistics, Old Dominion University, Norfolk, VA, USA\\
\small \textsuperscript{4}LightSpin Consulting LLC, Arlington, MA, USA\\
\small \textsuperscript{5}Department of Materials Science and Engineering, Cornell University, Ithaca, NY, USA}
\date{}

\begin{document}
\maketitle

\begin{abstract}
Modern position-sensitive PET detectors increasingly provide continuous estimates of photon interaction position, together with event-specific timing, energy, and depth-of-interaction (DOI) information. Taking full advantage of these measurements requires a projection model that preserves the measured coordinates and accounts for their uncertainties. We introduce the Spatial Probability Localization and Transformation (SPLAT) model, a continuous-domain, event-specific projection model for time-of-flight (TOF) list-mode PET reconstruction in which each detected coincidence is represented by a three-dimensional spatial response kernel in image space. The kernel is incorporated directly into consistent forward and adjoint projection operators without requiring detector-coordinate binning. Although the formulation permits general response distributions, we develop an anisotropic Gaussian implementation in which the kernel center is determined from the measured interaction coordinates and timing difference, and its full covariance is constructed by propagating detector measurement uncertainties through the event geometry. The model incorporates DOI, photon non-collinearity, and energy-dependent, correlated, and depth-dependent detector responses. We evaluate SPLAT using GATE Monte Carlo simulations with known annihilation positions, comparing continuous and virtually segmented representations of the same accepted coincidence events. Event-localization and Gaussian-reference diagnostics assess the accuracy of the predicted responses, while point-source and brain-phantom reconstructions demonstrate the consequences for image formation.  The Gaussian response also provides a natural basis for estimating information-equivalent count gain, an approximate measure of the relative spatial information provided by detected coincidences that accounts for their localization uncertainty and detection efficiency.  The results support the accuracy and utility of the event-specific formulation, particularly the incorporation of measured DOI, and an optimized GPU implementation demonstrates its computational feasibility. SPLAT provides a flexible framework for incorporating detailed detector measurements and their uncertainties directly into PET reconstruction.
\end{abstract}


\section{Introduction}

The system model in positron emission tomography (PET) describes the relationship between the radiotracer activity distribution and detected coincidence events.   Conventional PET models commonly represent detector coordinates by crystal or detector-element indices, with measurements organized as discrete lines of response (LORs), histogrammed projections, or list-mode events assigned to detector-element centers.   This representation is well suited to pixelated scintillator arrays and has supported efficient analytical and iterative reconstruction methods \cite{parra1998list, reader2002onepass, qi2006iterative, daubewitherspoon2012comparison}.

Advances in monolithic scintillators, finely pixelated detectors, and depth-of-interaction (DOI) estimation increasingly permit photon interaction positions to be measured continuously in three dimensions, together with deposited energy and detection time \cite{maas2009monolithic, vandam2011doi, borghi2016towards, gonzalez2021evolution}.   These measurements provide spatial information and event-specific localization uncertainties that are not fully represented by a detector-element index.   Nevertheless, continuous position estimates are often assigned to virtual crystals or other discrete coordinates before reconstruction.   Although this simplifies data handling, it can discard information within detector bins \cite{cabello2012monolithic, Clarkson:20}.   The resulting information loss depends on the bin dimensions relative to the intrinsic detector resolution and on the imaging task.

List-mode reconstruction avoids histogramming coincidences into conventional sinograms and can retain individual event attributes \cite{parra1998list, reader2002onepass, daubewitherspoon2012comparison}.   However, list-mode processing does not necessarily preserve continuous detector coordinates or their associated uncertainties.   Existing resolution models address detector blurring through measured or Monte Carlo-derived system matrices, spatially varying response functions, image-space point-spread functions, and explicit DOI and parallax corrections \cite{rogers1995doi, alessio2006modeling, panin2006fully, moehrs2008multiray, cloquet2010nongaussian, rahmim2013resolution,zhang2016efficient}.

Several earlier studies have incorporated continuous detector coordinates and event-specific response information.   Staelens \textit{et al.}\ modeled Gaussian interaction-position uncertainties \cite{Staelens_2004}, while Iriarte \textit{et al.}\ developed analytic response models incorporating DOI and depth-dependent positioning uncertainty \cite{Iriarte_2009}.   Caucci \textit{et al.}\ used probability distributions of three-dimensional interaction-position estimates in list-mode reconstruction \cite{Caucci_2010}.   Cabello \textit{et al.}\ retained continuous interaction coordinates and compared them with artificially discretized measurements \cite{cabello2012monolithic}.   Other approaches have used elliptical Gaussian tubes of response \cite{ortuno2011projector, sportelli2013elliptical}, anisotropic or asymmetric LOR probability densities \cite{yao2012lorpdf}, spatially varying event-response distributions \cite{bickell2016spatially}, and event-specific uncertainty estimates for reconstruction weighting \cite{thull2026uncertainty}.   These studies provide important precedents for continuous-coordinate and event-specific PET projection models.

Building on these developments, we introduce the Spatial Probability Localization and Transformation (SPLAT) model, an event-specific, continuous-domain projection model for time-of-flight (TOF) list-mode PET reconstruction.   In SPLAT, each detected coincidence defines a three-dimensional probability kernel in image space, constructed from the measured event attributes and their uncertainties.   The formulation preserves continuous detector coordinates independently of the image discretization and uses the same event-response kernel in the integral forward and adjoint back-projection operators.   The name SPLAT reflects the interpretation of back-projection as the accumulation of spatially distributed event responses, analogous to splatting in computer graphics \cite{zwicker2001surface, kerbl2023gaussian}.   The formulation is not restricted to a particular choice of response kernel.

In the Gaussian implementation examined here, the kernel center is determined from the measured detector endpoints and their detection-time difference, with corrections for correlated measurement errors.   The full three-dimensional covariance is constructed by propagating detector-local uncertainties through the event-specific TOF geometry using Jacobians.   This provides a unified treatment of detector-position uncertainty, DOI, coincidence timing resolution, photon non-collinearity, and energy-dependent, correlated, and depth-dependent detector responses \cite{shibuya2007acollinearity, moses2011limits, vandam2011doi, vandam2013sub200, tabacchini2015time, borghi2016towards}.   Each coincidence therefore has a spatial response whose center, extent, and orientation reflect its measured attributes and detector uncertainties.

We evaluate SPLAT using GATE Monte Carlo simulations of a continuous cylindrical LSO detector and matched virtually segmented representations of the same coincidence data.  A hierarchy of projection models isolates the contributions of continuous detector coordinates, measured DOI, energy correlations, and depth-dependent spatial resolution.  Using known annihilation positions, we evaluate event-localization accuracy independently of positron range and compare the resulting models through point-source and brain-phantom reconstructions.  The event-specific Gaussian responses also provide a natural basis for estimating information-equivalent count gain, which combines accepted-count efficiency with the spatial concentration of the event-response kernels to compare the localization information provided by different projection models.  Finally, we examine the computational performance of the GPU implementation.

\section{Methods}

We first define the measured coincidence coordinates and formulate the continuous-domain forward and adjoint projection operators. We then introduce the event-specific Gaussian kernel and derive its covariance from detector measurement uncertainties, including DOI, TOF, photon non-collinearity, and energy-dependent detector response. Additional system-response components, image discretization, and the reconstruction algorithm are described before presenting the Monte Carlo simulation and detector digitization methods. Finally, we evaluate the projection models using known annihilation positions from the simulations, providing a direct physical reference for assessing event-localization accuracy. Matched continuous and virtually segmented representations allow us to isolate the effects of detector-coordinate quantization and the additional response-model components. We conclude with information-equivalent count comparisons and reconstruction-based evaluations.

\subsection{Measurement and event coordinates}
\label{sec:measurementCoordinates}

A list-mode TOF coincidence can be represented by
\begin{equation}
    u
    =
    (\mathbf{y}_1,E_1\,;\,
     \mathbf{y}_2,E_2\,;\,\Delta \tau),
    \qquad
    \mathbf{y}_j\in\mathcal{D}\subset\mathbb{R}^3,
    \qquad \Delta \tau \in \mathbb{R}_+
\label{eq:rawEvent}
\end{equation}
where \(\mathbf{y}_j\) and \(E_j\) are the measured interaction-position estimate and energy for detector side \(j\), and $\Delta \tau$ is the measured timing difference \cite{parra1998list, reader2002onepass, daubewitherspoon2012comparison}.   The detector indexes are ordered by measured arrival time.   For a continuous position-sensitive detector, \(\mathbf{y}_j\) is retained as a continuous coordinate \cite{maas2009monolithic, vandam2011doi, borghi2016towards, cabello2012monolithic}.

For each event the measured LOR geometry and TOF-localized event position are
\begin{equation}
\begin{gathered}
    \ell:=\|\mathbf y_2-\mathbf y_1\|_2,\qquad
    \mathbf e:=\frac{\mathbf y_2-\mathbf y_1}{\ell}, \qquad
    r:=\frac{\ell-c\Delta\tau}{2}, \\
    \eta:=\frac{c\Delta\tau}{2},\qquad \text{and} \qquad
    \mathbf{z}:=\frac{\mathbf{y}_1+\mathbf{y}_2}{2}
                 -\eta\mathbf{e}.
\end{gathered}
\label{eq:eventGeometry}
\end{equation}
Here, \(\ell\) is the measured LOR length, and \(\mathbf e\) is the unit vector directed from the earlier-arrival endpoint \(\mathbf y_1\) to the later-arrival endpoint \(\mathbf y_2\).   The quantity \(\eta\) converts the time difference into a displacement from the LOR midpoint, and \(r\) is the signed distance along the oriented LOR from \(\mathbf y_1\) to the TOF-localized event position \(\mathbf z\).   In DIRECT reconstruction, TOF-localized events are accumulated into view-grouped histo-images \cite{Matej_2009,daubewitherspoon2012comparison}.   Because \(\Delta\tau\geq0\), the displacement from the LOR midpoint is toward the earlier-arrival endpoint \cite{conti2009state}.   

At each endpoint, detector-position uncertainty is expressed in a local orthonormal detector frame.   Let
\begin{equation}
    \mathbf R_j
      =
      [\,\mathbf t_j\ \mathbf a_j\ \mathbf n_j\,],
    \qquad
    \mathbf R_j^{\mathsf T}\mathbf R_j=\mathbf I_3,
    \qquad j\in\{1,2\},
\label{eq:detectorFrame}
\end{equation}
where \(\mathbf t_j\), \(\mathbf a_j\), and \(\mathbf n_j\) are unit vectors in the detector-tangential, detector-axial, and detector-normal (DOI) directions, expressed in the global image-space coordinate system.   Thus, \(\mathbf R_j\) is the rotation matrix that maps a detector-local vector into image-space coordinates.   

\subsection{Continuous-domain PET forward and adjoint model}
\label{sec:continuousForwardModel}


Let \(\Omega\subset\mathbb{R}^3\) be the image domain and \(\mathcal M\) the continuous
event domain.   An event descriptor \(m\in\mathcal M\) collects the measured attributes
in~\eqref{eq:rawEvent} and the derived geometry in~\eqref{eq:eventGeometry}.
For a nonnegative source density \(f\), the mean detected-event density is
\begin{equation}
\begin{aligned}
    g(m;f)
      &=
      (\mathcal Kf)(m)+\gamma(m),\\
    (\mathcal Kf)(m)
      &:=
      \int_{\Omega}
      K(m,\mathbf x)f(\mathbf x)\,d\mathbf x,
      \qquad m\in\mathcal M,
\end{aligned}
\label{eq:continuousForwardModel}
\end{equation}
where \(K(m,\mathbf x)\geq 0\) is the system-response density for each event \(m\), and \(\gamma(m)\geq 0\) is the additive background density.  The event coordinates remain continuous independently of the finite image representation introduced later.

The adjoint operator and image-space sensitivity are
\begin{equation}
    (\mathcal K^{*}h)(\mathbf x)
      :=
      \int_{\mathcal M}
      K(m,\mathbf x)h(m)\,dm,
    \qquad
    s(\mathbf x)
      :=
      (\mathcal K^{*}\mathbf 1)(\mathbf x),
\label{eq:continuousAdjointSensitivity}
\end{equation}
with
\begin{equation}
    \langle\mathcal Kf,h\rangle
      =
      \langle f,\mathcal K^{*}h\rangle.
\label{eq:continuousAdjointIdentity}
\end{equation}
Thus, the same response kernel defines forward projection and back-projection \cite{parra1998list, reader2002onepass, qi2006iterative}.

\subsubsection{List-mode Poisson data model and likelihood}
\label{sec:listModeLikelihood}

The accepted events are modeled as an inhomogeneous Poisson point process with intensity \(g(m;f)\) \cite{nichols2002spatiotemporal, snyder2007parameter, snyder2012random}.  The expected number of events is
\begin{equation}
    \overline N(f)
      =
      \int_{\mathcal M}g(m;f)\,dm
      =
      \int_{\Omega}s(\mathbf x)f(\mathbf x)\,d\mathbf x
      +\overline\gamma,
    \qquad
    \overline\gamma
      :=
      \int_{\mathcal M}\gamma(m)\,dm .
\label{eq:expectedTotalEvents}
\end{equation}

For observed events \(\mathcal Y=\{m_n\}_{n=1}^{N}\), the list-mode likelihood is
\begin{equation}
    L(f;\mathcal Y)
      \propto
      \exp[-\overline N(f)]
      \prod_{n=1}^{N}
      \left[
        (\mathcal Kf)(m_n)+\gamma(m_n)
      \right]
\label{eq:listModePoissonLikelihood}
\end{equation}
\cite{barrett1997list,parra1998list}.  If the background model is independent of \(f\), the corresponding negative log-likelihood is 
\begin{equation}
    \ell(f)
      =
      \int_{\Omega}s(\mathbf x)f(\mathbf x)\,d\mathbf x
      -
      \sum_{n=1}^{N}
      \log
      \left[
        (\mathcal Kf)(m_n)+\gamma(m_n)
      \right]
      +\mathrm{const.}
\label{eq:listModeNegativeLogLikelihoodContinuous}
\end{equation}

Penalized-likelihood reconstruction solves
\begin{equation}
    \widehat f
      =
      \operatorname*{arg\,min}_{f\geq0}
      \left\{
        \ell(f)+\lambda\Phi(f)
      \right\},
\label{eq:continuousPenalizedLikelihood}
\end{equation}
Here, \(\Phi\) is a regularization functional \cite{qi2006iterative}.  Its discretization and numerical solution are described in section~\ref{sec:numericalReconstruction}.

\subsection{Event-specific Gaussian image-space kernel}
\label{sec:eventSpecificKernel}

\begin{figure}[htb]
    \centering
    \includegraphics[width=0.6\linewidth]{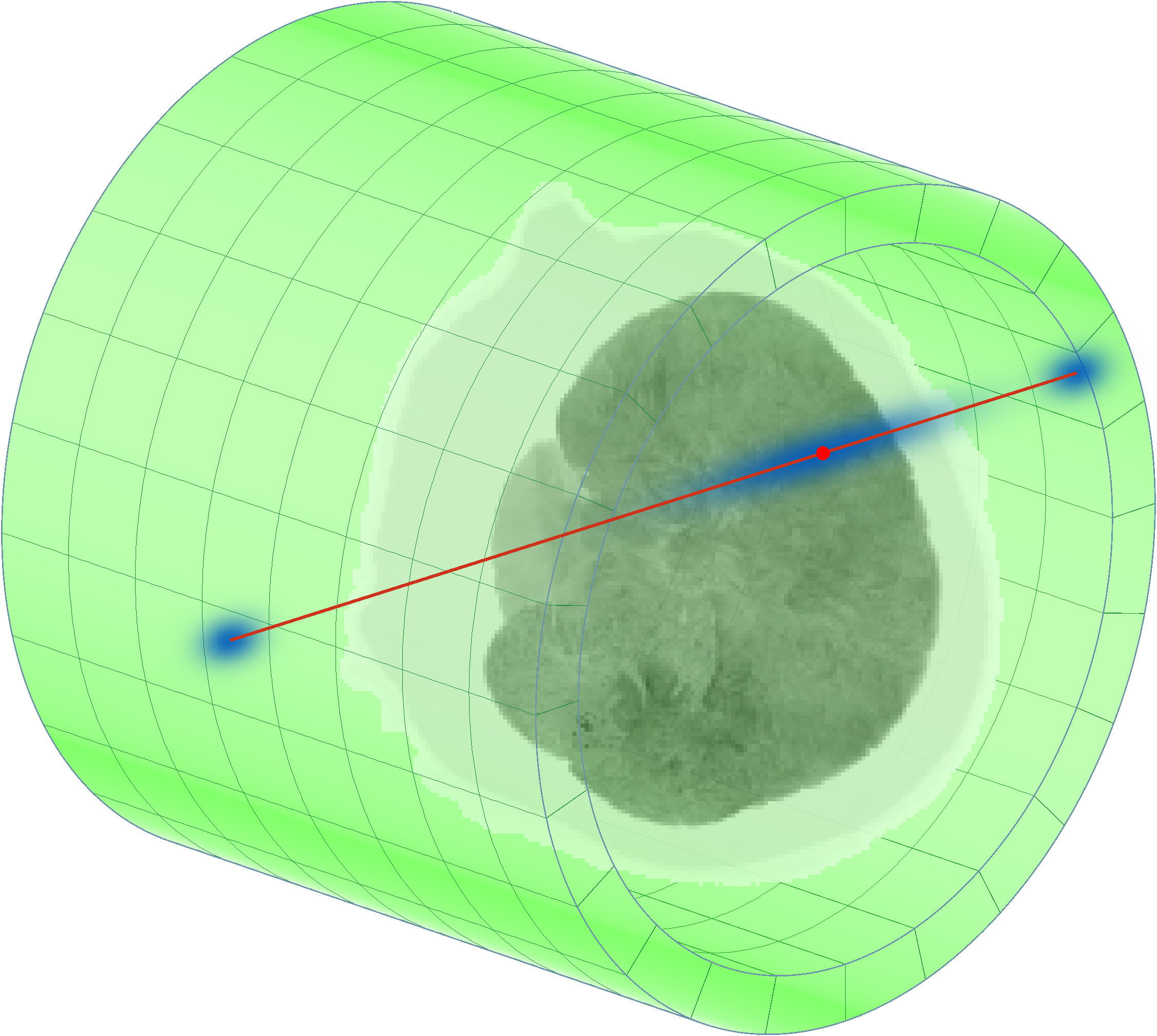}
    \caption{Illustration of the three-dimensional, event-specific
    Gaussian kernel used in the SPLAT projection model.}
    \label{fig:2D_SPLAT}
\end{figure}

Continuous detector readout provides event-specific estimates of interaction position, time, energy, and DOI, with uncertainties that may be anisotropic and correlated \cite{maas2009monolithic,vandam2011doi,tabacchini2015time, borghi2016towards}.  When these uncertainties are approximately Gaussian, their propagation through the TOF localization map produces an approximately Gaussian distribution in image space.  A three-dimensional Gaussian therefore provides a compact description of the event's localization uncertainty (Fig.~\ref{fig:2D_SPLAT}).  Elliptical Gaussian tubes of response have previously been used in PET reconstruction \cite{ortuno2011projector,sportelli2013elliptical}, alongside other anisotropic and spatially varying event-response models \cite{yao2012lorpdf,bickell2016spatially}.

For event \(m\), the Gaussian kernel is
\begin{equation}
    \kappa_{\mathrm G}(m,\mathbf x)
      =
      \frac{
        \exp\!\left[
          -\tfrac12
          \bigl(\mathbf x-\boldsymbol{\mu}(m)\bigr)^{\mathsf T}
          \boldsymbol{\Sigma}(m)^{-1}
          \bigl(\mathbf x-\boldsymbol{\mu}(m)\bigr)
        \right]
      }{
        (2\pi)^{3/2}
        \sqrt{\det\boldsymbol{\Sigma}(m)}
      },
\label{eq:eventGaussianKernel}
\end{equation}
where \(\boldsymbol{\mu}(m)\) and \(\boldsymbol{\Sigma}(m)\) are the event-specific mean
and covariance, and \(\boldsymbol{\Sigma}(m)\) is symmetric positive definite.  The kernel is normalized over \(\mathbb R^3\).  The mean is the TOF-localized position \(\mathbf z(m)\), unless an energy-dependent correction is applied.

The eigen-decomposition
\begin{equation}
    \boldsymbol{\Sigma}(m)
      =
      \mathbf U(m)
      \operatorname{diag}
      \left(
        \sigma_1^2(m),
        \sigma_2^2(m),
        \sigma_3^2(m)
      \right)
      \mathbf U(m)^{\mathsf T}
\label{eq:kernelPrincipalAxes}
\end{equation}
defines the principal directions and standard deviations of the Gaussian response.  Both the mean and covariance vary between events according to the measured geometry and detector response.  The covariance is constructed in section~\ref{sec:jacobianCovariance}, and the remaining system-response factors are described in section~\ref{sec:kernelFactorization}.

The SPLAT forward- and back-projection operators have a simple interpretation in terms of event-specific image-space responses (Fig.~\ref{fig:splat-pseudocode}).  Each coincidence defines a Gaussian splat, specified by its center \(\boldsymbol{\mu}_m\) and covariance \(\boldsymbol{\Sigma}_m\).  Forward-projection computes the splat's contribution to the measured event, while back-projection accumulates these splats in image space.  The figure illustrates these operations using MATLAB-like pseudocode, with Gaussian integration over image voxels represented by \texttt{gaussFilter}.

\begin{figure*}[htb]
\centering
\begin{minipage}[t]{0.4965\textwidth}
\vspace{0pt}
\noindent
\colorbox{blue!10}{%
    \parbox[c][2em][c]{\dimexpr\linewidth-2\fboxsep\relax}{%
        \centering\footnotesize
        \strut
        \splatHeader{blue}{Forward Projection}{ 
        (\mathcal K f)_m =\langle\kappa_m,f\rangle }
        \strut
    }%
}
\vspace{1pt}
\begin{splatCode}
fProj = zeros(nEvents,1);
for m = 1:nEvents
    I_m = gaussFilter(delta(mu_m),Sigma_m);
    fProj(m) = sum(image.*I_m,'all');
end
\end{splatCode}

\end{minipage}
\hfill
\begin{minipage}[t]{0.4965\textwidth}
\vspace{0pt}
\noindent
\colorbox{red!10}{%
    \parbox[c][2em][c]{\dimexpr\linewidth-2\fboxsep\relax}{%
        \centering\footnotesize
        \strut
        \splatHeader{red}{Back-Projection}{
        \mathcal K^\ast h =\sum_m h_m\kappa_m }
        \strut
    }%
}
\vspace{1pt}
\begin{splatCode}
bProj = zeros(size(image));
for m = 1:nEvents
    I_m = gaussFilter(delta(mu_m),Sigma_m);
    bProj = bProj + eventValue(m)*I_m;
end
\end{splatCode}
\end{minipage}
\caption{
Conceptual SPLAT forward- and back-projection operators expressed in MATLAB-like pseudocode.  Both operators use the same discretized kernel.  Multiplicative correction factors are omitted for clarity.
}
\label{fig:splat-pseudocode}
\end{figure*}

\subsection{Event-specific covariance construction}
\label{sec:jacobianCovariance}

The mapping from the measured detector attributes to the TOF-localized image position in~\eqref{eq:eventGeometry} is locally linearized, yielding event-specific Jacobians.  These Jacobians propagate detector-local timing and position uncertainties into scanner coordinates.  The error of the TOF-localized position relative to the true annihilation position \(\mathbf z_{\mathrm{true}}\) has two independent parts,
\begin{equation}
    \mathbf z-\mathbf z_{\mathrm{true}}
      =
      \delta\mathbf z+\delta\mathbf z_{\mathrm{NC}},
\label{eq:errorDecomposition}
\end{equation}
where \(\delta\mathbf z\) is the error propagated from the detector measurements and \(\delta\mathbf z_{\mathrm{NC}}\) is the transverse offset caused by photon non-collinearity, so that the event covariance is
\begin{equation}
    \boldsymbol{\Sigma}(m)
      =
      \boldsymbol{\Sigma}_{\mathrm{det}}(m)
      +
      \boldsymbol{\Sigma}_{\mathrm{NC}}(m),
\label{eq:eventCovariancePropagation}
\end{equation}
where \(\boldsymbol{\Sigma}_{\mathrm{det}}:=\operatorname{Cov}(\delta\mathbf z)\) collects the timing, tangential, axial, and DOI uncertainties of the two photon detections, mapped through the event geometry, and \(\boldsymbol{\Sigma}_{\mathrm{NC}}:=\operatorname{Cov}(\delta\mathbf z_{\mathrm{NC}})\).  The detector uncertainties may also depend on measured energy, reflecting the relationship between scintillation light yield and the statistical precision of the position and timing measurements.  Correlations among these measurements may also be incorporated through their joint response with energy.

\subsubsection{Event covariance propagation}

The projection matrix onto the plane perpendicular to the measured LOR direction \(\mathbf e\) is
\begin{equation}
    \mathbf P_{\perp}
      :=
      \mathbf I_3-\mathbf e\mathbf e^{\mathsf T}.
\label{eq:transverseProjector}
\end{equation}
Without loss of generality we define \(\tau_1\) and \(\tau_2\) as the detection times of the two photons, so that the measured timing difference in~\eqref{eq:eventGeometry} is \(\Delta\tau:=\tau_2-\tau_1\).  From equation~\eqref{eq:eventGeometry}, the derivatives of the TOF-localized position with respect to the detection positions and times are
\begin{equation}
    \mathbf J_1
      :=\frac{\partial\mathbf z}{\partial\mathbf y_1}
      =
      \frac12\mathbf I_3
      +\frac{\eta}{\ell}\mathbf P_{\perp},
    \qquad
    \mathbf J_2
      :=\frac{\partial\mathbf z}{\partial\mathbf y_2}
      =
      \frac12\mathbf I_3
      -\frac{\eta}{\ell}\mathbf P_{\perp},
    \qquad
    \frac{\partial\mathbf z}{\partial\tau_1}
      =\frac{c}{2}\mathbf e,
    \qquad
    \frac{\partial\mathbf z}{\partial\tau_2}
      =-\frac{c}{2}\mathbf e .
\label{eq:eventCenterJacobians}
\end{equation}

Let \(\delta\mathbf y_j\) denote the error in the detection position \(\mathbf y_j\) and \(\delta\tau_j\) the error in the detection time \(\tau_j\), each defined as the measured value minus the ideal value.  Linearizing~\eqref{eq:eventGeometry} about the measured values gives the measurement part \(\delta\mathbf z\) of~\eqref{eq:errorDecomposition}, 
\begin{equation}
    \delta\mathbf z
      \approx
      \mathbf J_1\,\delta\mathbf y_1
      +\mathbf J_2\,\delta\mathbf y_2
      +\frac{c}{2}\mathbf e\,\delta\tau_1
      -\frac{c}{2}\mathbf e\,\delta\tau_2 .
\label{eq:firstOrderDisplacement}
\end{equation}
The detector resolution is specified along the detector's own axes, so the position error at side \(j\) is written in the detector-local frame of~\eqref{eq:detectorFrame} as
\begin{equation}
    \mathbf s_j
      :=
      \begin{bmatrix}
        s_{\mathrm{tan},j}\\
        s_{\mathrm{ax},j}\\
        s_{\mathrm{doi},j}
      \end{bmatrix},
    \qquad
    \delta\mathbf y_j=\mathbf R_j\mathbf s_j ,
\label{eq:localPositionError}
\end{equation}
with components along the tangential, axial, and DOI axes.  For each photon detection \(j\in\{1,2\}\), the timing and position errors are collected into
\begin{equation}
    \mathbf q_j
      :=
      \begin{bmatrix}
        \delta\tau_j\\
        \mathbf s_j
      \end{bmatrix},
    \qquad
    \mathbf C_{q_j}
      :=
      \operatorname{Cov}(\mathbf q_j).
\label{eq:localResponseVector}
\end{equation}
Combining the timing and position contributions yields the \(3\times4\) detector-response Jacobians
\begin{equation}
    \mathbf H_1
      =
      \begin{bmatrix}
        \frac{c}{2}\mathbf e & \mathbf J_1\mathbf R_1
      \end{bmatrix},
    \qquad
    \mathbf H_2
      =
      \begin{bmatrix}
        -\frac{c}{2}\mathbf e & \mathbf J_2\mathbf R_2
      \end{bmatrix},
\label{eq:sideResponseMatrices}
\end{equation}
so that \eqref{eq:firstOrderDisplacement} becomes
\begin{equation}
    \delta\mathbf z
      \approx
      \mathbf H_1\mathbf q_1+\mathbf H_2\mathbf q_2 .
\label{eq:eventPositionPerturbation}
\end{equation}
Assuming that the two detections have independent measurement errors, so that cross terms between them vanish, \(\operatorname{Cov}(\mathbf A\mathbf x)=\mathbf A\operatorname{Cov}(\mathbf x)\mathbf A^{\mathsf T}\) applied to each term gives
\begin{equation}
\boxed{\;
    \boldsymbol{\Sigma}_{\mathrm{det}}(m)
      =
      \sum_{j=1}^{2}
      \mathbf H_j\mathbf C_{q_j}\mathbf H_j^{\mathsf T},
      \;}
\label{eq:detectorCovariance}
\end{equation}
where \(\mathbf C_{q_j}\) is specified in Sections~\ref{sec:doiUncertainty} and~\ref{sec:tofUncertainty}.

Photon non-collinearity introduces an additional angular uncertainty, which displaces the event perpendicular to the LOR by an amount that depends on the position along the LOR.  We define the geometric factor
\begin{equation}
    r_{\mathrm c}
      :=
      \min\{\max(r,0),\ell\},
    \qquad
    g_{\mathrm{NC}}(r,\ell)
      :=
      \frac{r_{\mathrm c}(\ell-r_{\mathrm c})}{\ell},
\label{eq:noncollinearityGeometry}
\end{equation}
where \(r_{\mathrm c}\) restricts the event coordinate to the detector interval.  The angular uncertainty is modeled as isotropic and zero mean, with per-axis standard deviation \(\sigma_\alpha\) given in section~\ref{sec:nonCollinearity}.  Under the small-angle approximation, the resulting transverse displacement \(\delta\mathbf z_{\mathrm{NC}}\) is isotropic with per-axis standard deviation \(g_{\mathrm{NC}}(r,\ell)\,\sigma_\alpha\), so its covariance is   
\begin{equation}
\boxed{\;
    \boldsymbol{\Sigma}_{\mathrm{NC}}(m)
      =                                                      
      \left[g_{\mathrm{NC}}(r,\ell)\,\sigma_\alpha\right]^{2}\mathbf P_{\perp}.
      \;}
\label{eq:noncollinearityCovariance}
\end{equation}
Equations~\eqref{eq:detectorCovariance} and~\eqref{eq:noncollinearityCovariance} give the two terms of~\eqref{eq:eventCovariancePropagation}.  

\subsubsection{Detector-position and DOI uncertainty}
\label{sec:doiUncertainty}

The covariance of the local position error \(\mathbf s_j\) in~\eqref{eq:localPositionError} is \(\mathbf C_{s_j}:=\operatorname{Cov}(\mathbf s_j)\).  For independent components,
\begin{equation}
    \mathbf C_{s_j}
      =
      \operatorname{diag}
      \left(
        \sigma_{\mathrm{tan},j}^{2},
        \sigma_{\mathrm{ax},j}^{2},
        \sigma_{\mathrm{doi},j}^{2}
      \right),
\label{eq:detectorPositionCovariance}
\end{equation}
where the diagonal elements are the variances in the detector-tangential, detector-axial, and detector-normal (DOI) directions.  A full covariance may be used when these components are correlated.

Measured DOI determines the detector-normal coordinate of \(\mathbf y_j\), whereas \(\sigma_{\mathrm{doi},j}\) describes uncertainty in that estimate.  The spatial resolutions may depend on interaction depth, energy, incidence geometry, and detector position \cite{maas2009monolithic, vandam2011doi, tabacchini2015time, borghi2016towards}.

If timing and position errors are independent, the local measurement covariance in~\eqref{eq:localResponseVector} is
\begin{equation}
    \mathbf C_{q_j}
      =
      \begin{bmatrix}
        \sigma_{\tau,j}^{2} & \mathbf 0^{\mathsf T}\\
        \mathbf 0 & \mathbf C_{s_j}
      \end{bmatrix},
\label{eq:localMeasurementCovariance}
\end{equation}
where \(\sigma_{\tau,j}^{2}:=\operatorname{Var}(\delta\tau_j)\) is given in Section~\ref{sec:tofUncertainty}.

\subsubsection{TOF uncertainty}
\label{sec:tofUncertainty}

For independent
detection-time errors, the timing contribution to \(\boldsymbol{\Sigma}_{\mathrm{det}}\) in equation~\eqref{eq:detectorCovariance} is
\begin{equation}
    \boldsymbol{\Sigma}_{\mathrm{TOF}}
      =
      \frac{c^2}{4}
      \left(
        \sigma_{\tau,1}^{2}
        +\sigma_{\tau,2}^{2}
      \right)
      \mathbf e\mathbf e^{\mathsf T}
      =
      \sigma_{\parallel,\mathrm{TOF}}^2
      \mathbf e\mathbf e^{\mathsf T},
\label{eq:tofCovariance}
\end{equation}
where the TOF localization standard deviation is
\begin{equation}
    \sigma_{\parallel,\mathrm{TOF}}
      =
      \frac{c\sigma_{\Delta\tau}}{2},
    \qquad
    \sigma_{\Delta\tau}^{2}
      =
      \sigma_{\tau,1}^{2}+\sigma_{\tau,2}^{2},
    \qquad
    \sigma_{\Delta\tau}
      =
      \frac{\mathrm{CTR}_{\mathrm{FWHM}}}
           {2\sqrt{2\ln2}} .
\label{eq:tofSpatialResolution}
\end{equation}
When timing and position errors are correlated, their joint contribution is retained in \(\mathbf C_{q_j}\) \cite{conti2009state, vandam2013sub200}.

\subsubsection{Photon non-collinearity}
\label{sec:nonCollinearity}

The angular uncertainty in equation~\eqref{eq:noncollinearityCovariance} has standard deviation
\begin{equation}
    \sigma_{\alpha}
      =
      \frac{\alpha_{\mathrm{FWHM}}}
           {2\sqrt{2\ln2}} .
\label{eq:noncollinearityAngularCovariance}
\end{equation}
The transverse uncertainty is greatest at the LOR midpoint, where
\[
    \mathrm{FWHM}_{\perp,\mathrm{NC}}
      \simeq
      \frac{\ell}{4}\alpha_{\mathrm{FWHM}},
\]
and decreases toward either detector
\cite{shibuya2007acollinearity,moses2011limits}.

\subsubsection{Energy-dependent and correlated detector response}

The measured energy provides additional information about the timing and position uncertainties of each photon detection.  Higher deposited energy generally produces more scintillation photons, improving the precision of these measurements.  In addition, errors in the measured energy, time, and position may be correlated because they arise from the same scintillation signal.

For each photon detection \(j\), let
\[
    \delta E_j:=E_j-E_{\mathrm{ref}},
\]
where \(E_{\mathrm{ref}}\) is the reference energy.  We represent the joint measurement uncertainties by
\begin{equation}
    \boldsymbol{\epsilon}_j
      :=
      \begin{bmatrix}
        \delta E_j\\
        \mathbf q_j
      \end{bmatrix}
      \sim
      \mathcal N
      \left(
        \mathbf 0,
        \begin{bmatrix}
          \sigma_{E,j}^{2}
            & \mathbf c_{qE,j}^{\mathsf T}\\
          \mathbf c_{qE,j}
            & \mathbf C_{q_j}
        \end{bmatrix}
      \right),
\label{eq:jointEnergyResponse}
\end{equation}
where
\[
    \mathbf c_{qE,j}
      =
      \operatorname{Cov}(\mathbf q_j,\delta E_j)
\]
describes correlations between energy and the timing and position errors.  Equivalently, the joint covariance can be expressed as \(\mathbf D_j\boldsymbol{\Gamma}_j\mathbf D_j\), where \(\mathbf D_j\) contains the individual measurement standard deviations and \(\boldsymbol{\Gamma}_j\) is their correlation matrix.

Using the measured energy, the expected measurement errors and their covariance become
\begin{equation}
\begin{aligned}
    \boldsymbol{\mu}_{q|E,j}
      &=
      \frac{\mathbf c_{qE,j}}{\sigma_{E,j}^{2}}
      \,\delta E_j,\\
    \mathbf C_{q|E,j}
      &=
      \mathbf C_{q_j}
      -
      \frac{
        \mathbf c_{qE,j}\mathbf c_{qE,j}^{\mathsf T}
      }{
        \sigma_{E,j}^{2}
      }.
\end{aligned}
\label{eq:conditionalEnergyResponse}
\end{equation}
Because the perturbations are defined as measured minus ideal, the resulting correction to the Gaussian kernel center is
\begin{equation}
\boxed{
    \boldsymbol{\mu}(m)
      =
      \mathbf z
      -
      \sum_{j=1}^{2}
      \mathbf H_j\boldsymbol{\mu}_{q|E,j}.
      }
\label{eq:conditionalKernelMean}
\end{equation}
The corresponding detector and total event covariances are
\begin{equation}
\boxed{
\begin{aligned}
    \boldsymbol{\Sigma}_{\mathrm{det}}(m)
      &=
      \sum_{j=1}^{2}
      \mathbf H_j
      \mathbf C_{q|E,j}
      \mathbf H_j^{\mathsf T}.
\end{aligned}
}
\label{eq:conditionalKernelCovariance}
\end{equation}

When correlations with measured energy are neglected, \(\boldsymbol{\mu}(m)=\mathbf z\), and the original \(\mathbf C_{q_j}\) is used.  The individual timing and position uncertainties may still depend on measured energy, interaction depth, and detector position \cite{vandam2011doi,vandam2013sub200, tabacchini2015time,borghi2016towards}.

\subsection{Additional system-response components}
\label{sec:kernelFactorization}

\subsubsection{Multiplicative correction factors}

The Gaussian kernel describes the spatial distribution associated with each coincidence.  Attenuation, detector efficiency, and deadtime are included through multiplicative factors,
\begin{equation}
    K(m,\mathbf x)
      =
      w(m)\,\kappa_{\mathrm G}(m,\mathbf x),
    \qquad
    w(m)
      :=
      a_{\mu}(m)\,
      a_{\mathrm{norm}}(m)\,
      a_{\mathrm{DT}}(m),
\label{eq:systemKernelFactorization}
\end{equation}
where \(a_{\mu}\), \(a_{\mathrm{norm}}\), and \(a_{\mathrm{DT}}\) represent attenuation, detector efficiency, and deadtime, respectively \cite{jin2013list}.  Other multiplicative acquisition factors may be included in \(w\), while random and scattered coincidences are represented by the additive term \(\gamma\) \cite{qi2006iterative}.

For an unscattered, collinear photon pair, attenuation is
\begin{equation}
    a_{\mu}(m)
      =
      \exp\!\left[
        -\int_{[\mathbf y_1,\mathbf y_2]}
        \mu(\mathbf r)\,d\ell
      \right],
\label{eq:attenuationFactor}
\end{equation}
which is independent of the annihilation position along the LOR.  Detector efficiency is estimated from the 511-keV photoelectric interaction probabilities associated with the photon path lengths through the detector.  The coincidence efficiency is the product of the two photon-detection probabilities.

\subsubsection{Positron range}
\label{sec:positronRangeOperator}

The Gaussian event kernel describes uncertainty in the annihilation position, whereas the activity distribution describes the location of radioactive decay.  Positron range is therefore modeled separately, through an operator that maps the decay distribution to the annihilation distribution,
\begin{align}
    f_{\mathrm a}(\mathbf x_{\mathrm a})
      &=
      (\mathcal P_{\mathrm{pr}}f_{\mathrm d})
      (\mathbf x_{\mathrm a})
      :=
      \int_{\Omega}
      p_{\mathrm{pr}}
      (\mathbf x_{\mathrm a}\mid\mathbf x_{\mathrm d})
      f_{\mathrm d}(\mathbf x_{\mathrm d})
      \,d\mathbf x_{\mathrm d},
    \label{eq:positronRangeOperator}\\
    g(m;f_{\mathrm d})
      &=
      \bigl(
        \mathcal K_{\mathrm a}
        \mathcal P_{\mathrm{pr}}f_{\mathrm d}
      \bigr)(m)
      +\gamma(m),
    \label{eq:completeForwardWithPR}
\end{align}
where \(f_{\mathrm d}\) and \(f_{\mathrm a}\) are the decay- and annihilation-position distributions, \(p_{\mathrm{pr}}\) describes the radionuclide- and medium-dependent positron range, and \(\mathcal K_{\mathrm a}\) is the projection operator for the annihilation distribution \cite{levin1999positron,moses2011limits}.

Because positron transport occurs before annihilation, its uncertainty is not included in the event covariance \(\boldsymbol{\Sigma}(m)\).  When positron range is neglected, the source distribution \(f\) in section~\ref{sec:continuousForwardModel} is identified with \(f_{\mathrm a}\).

\subsection{Numerical realization and image reconstruction}
\label{sec:numericalReconstruction}

\subsubsection{Image representation and kernel discretization}
\label{sec:imageDiscretization}

The source distribution is approximated using image-basis functions,
\begin{equation}
    f_h(\mathbf x)
      =
      \sum_{j=1}^{d}f_j b_j(\mathbf x),
    \qquad
    f_j\geq0 ,
\label{eq:imageBasisRepresentation}
\end{equation}
where, \(b_j\) denotes the image-basis functions and \(f_j\) their nonnegative coefficients.  Separating a continuous integral model from its numerical approximation follows the integral-equation approach used in image restoration and emission tomography \cite{lu2010integral,jiang2018higher,tang2021integral}.  Piecewise-constant voxels are used in the reconstructions presented here.

For observed event \(m_n\), the corresponding system-matrix element and sensitivity coefficient are 
\begin{equation}
\begin{aligned}
    A_{nj}
      &:=
      \int_{\Omega}
      K(m_n,\mathbf x)b_j(\mathbf x)\,d\mathbf x
      =
      w(m_n)
      \int_{\Omega}
      \kappa_{\mathrm G}(m_n,\mathbf x)b_j(\mathbf x)\,d\mathbf x,\\
    B_j
      &:=
      \int_{\Omega}s(\mathbf x)b_j(\mathbf x)\,d\mathbf x .
\end{aligned}
\label{eq:discreteSystemMatrix}
\end{equation}
The kernel integral is evaluated by Gaussian quadrature over the support of \(b_j\),
\begin{equation}
    A_{nj}
      \simeq
      w(m_n)
      \sum_{q=1}^{Q}
      \omega_{jq}\,
      \kappa_{\mathrm G}(m_n,\mathbf x_{jq})
      b_j(\mathbf x_{jq}),
\label{eq:kernelQuadrature}
\end{equation}
where \(\mathbf x_{jq}\) and \(\omega_{jq}\) are the mapped quadrature points and weights.  The same coefficients are used for forward projection and back-projection,
\[
    (\mathbf A\mathbf f)_n=\sum_jA_{nj}f_j,
    \qquad
    (\mathbf A^{\mathsf T}\mathbf q)_j=\sum_nA_{nj}q_n .
\]
System-matrix rows are evaluated directly from the measured event attributes, without binning the detector coordinates, as in earlier continuous-coordinate list-mode implementations \cite{cabello2012monolithic}.

\subsubsection{Discrete penalized-likelihood objective}
\label{sec:discreteObjective}

Using the discrete image representation, the penalized-likelihood objective becomes
\begin{equation}
    \widehat{\mathbf f}
      =
      \operatorname*{arg\,min}_{\mathbf f\geq0}
      \left\{
        G(\mathbf f)
        +\lambda\Phi(\mathbf L\mathbf f)
      \right\},
\label{eq:discretePenalizedLikelihood}
\end{equation}
where \(\mathbf L\) is the selected regularization operator and
\begin{equation}
    G(\mathbf f)
      =
      \langle\mathbf B,\mathbf f\rangle
      -
      \left\langle
        \log
        \left(
          \mathbf A\mathbf f+\bar{\boldsymbol{\gamma}}
        \right),
        \mathbf 1
      \right\rangle .
\label{eq:discreteDataFidelity}
\end{equation}
Here \(\bar{\gamma}_n=\gamma(m_n)\), and the logarithm is applied componentwise.  The gradient of the smooth data term is
\begin{equation}
    \nabla G(\mathbf f)
      =
      \mathbf B
      -
      \mathbf A^{\mathsf T}
      \left( \frac{1}{\mathbf A\mathbf f+\bar{\boldsymbol{\gamma}}}
      \right) \,,
\label{eq:discreteLikelihoodGradient}
\end{equation}
where division is componentwise~\cite{barrett1997list,parra1998list,qi2006iterative}.

\subsubsection{Preconditioned reconstruction algorithm}
\label{sec:preconditionedAlgorithm}

Reconstruction uses a preconditioned primal--dual iteration following the Krasnosel'ski\u{\i}--Mann framework of \cite{lin2019krasnoselskii,wu2016reducing, jiang2018higher}.  The sensitivity-based diagonal preconditioner and relaxation schedule used here are specified below.  For image estimate \(\mathbf f^k\) and dual variable \(\mathbf b^k\), the updates are
\begin{equation}
\begin{aligned}
    \mathbf b^{k+1}
      &=
      \mu_k
      \left(
        \mathbf I
        -
        \operatorname{prox}_{(\lambda/\mu_k)\Phi}
      \right)
      \left(
        \mu_k^{-1}\mathbf b^k+\mathbf L\mathbf f^k
      \right),\\
    \mathbf f^{k+1}
      &=
      P_{+}
      \left[
        \mathbf f^k
        -
        \beta_k\mathbf S_k
        \left\{
          \nabla G(\mathbf f^k)
          +
          \mathbf L^{\mathsf T}
          \left(
            2\mathbf b^{k+1}-\mathbf b^k
          \right)
        \right\}
      \right],
\end{aligned}
\label{eq:preconditionedKma}
\end{equation}
where \(P_{+}\) denotes projection onto the nonnegative orthant.  On the positive-sensitivity image support, the diagonal preconditioner and step parameters are
\begin{equation}
    \mathbf S_k
      =
      \operatorname{diag}
      \left(
        \frac{\mathbf f^k}{\mathbf B}
      \right),
    \qquad
    \mu_k
      =
      \frac{1}{
        2\|\mathbf L\|_2^2\|\mathbf S_k\|_2
      },
    \qquad
    \beta_k
      =
      \frac{1}{1+a(k-1)},
\label{eq:preconditionerParameters}
\end{equation}
where \(a\geq0\) controls the decreasing relaxation schedule \cite{ahn2003globally,ross2017relaxed, lin2019krasnoselskii, guo2022fast}.

For ordered subsets, the event rows are partitioned once into \(n_{\mathrm{sub}}\) subsets and processed sequentially using
\begin{equation}
    \nabla G_i(\mathbf f)
      =
      \frac{\mathbf B}{n_{\mathrm{sub}}}
      -
      \mathbf A_i^{\mathsf T}
      \left( \frac{1}{\mathbf A_i\mathbf f + \bar{\boldsymbol{\gamma}}_i }
      \right).
\label{eq:subsetLikelihoodGradient}
\end{equation}
The corresponding preconditioner replaces \(\mathbf B\) by \(\mathbf B/n_{\mathrm{sub}}\).  Decreasing relaxation is used to reduce ordered-subset limit-cycle behavior \cite{ahn2003globally,ross2017relaxed}.

\subsection{Simulation data and model evaluation}
\label{sec:simulations}

\subsubsection{Monte Carlo simulation}
\label{sec:detectorSimulation}

GATE 9.4.1 was used to simulate \(^{18}\mathrm{F}\) positron decay and transport, annihilation-photon production and transport, and detector energy deposition \cite{jan2004gate,jan2011gate}.  GATE Actors recorded the decay and annihilation positions, allowing positron range to be distinguished from subsequent photon-transport and detector effects.  The scanner was modeled as a continuous cylindrical LSO shell with density \(7.4~\mathrm{g/ml}\), inner and outer radii of 125 and 145 mm, respectively, and an axial extent from \(-250\) to \(250\) mm.  GATE modeled the underlying particle interactions and energy deposition, while detector-readout uncertainties were introduced separately by the external digitizer described below.

\subsubsection{Detector digitization and event selection}
\label{sec:detectorDigitization}

To implement the required detector-response model, a custom post-processing digitizer was developed following the general approach used in GATE digitization.  Detector hits were converted into singles and subsequently into coincidences.  Spatiotemporally connected hits were grouped into clusters, with deposited energies summed and interaction positions, detection times, and DOI estimated using energy-weighted averages.  The analysis was restricted to coincidences in which both primary annihilation photons were absorbed solely through photoelectric interactions.

Four intrinsic detector-resolution cases were considered, with spatial FWHMs of \(1\), \(2\), \(3\), and \(4\) mm at a reference depth of 10 mm.  The measured attributes of each single were perturbed using a correlated Gaussian response,
\begin{equation}
    \begin{bmatrix}
        \delta E &
        \delta\tau &
        \delta s_{\mathrm{tan}} &
        \delta s_{\mathrm{ax}} &
        \delta s_{\mathrm{doi}}
    \end{bmatrix}^{\mathsf T}
    \sim
    \mathcal N
    \left(
        \mathbf 0,
        \mathbf D\boldsymbol{\Gamma}\mathbf D
    \right),
\label{eq:digitizerResponseDraw}
\end{equation}
where \(\mathbf D\) contains the event-specific marginal standard deviations and \(\boldsymbol{\Gamma}\) is the correlation matrix.  The digitizer used the underlying GATE interaction information to generate measurement perturbations, whereas the projection models used the resulting measured detector attributes to estimate each event's response.

Joint estimation of interaction position, DOI, energy, and timing is characteristic of monolithic scintillator readout \cite{schaart2009novel,vandam2011doi,vandam2013sub200, borghi2016towards}.  The response functions and numerical parameters were prescribed for this controlled simulation rather than fitted to a particular physical detector (Table~\ref{tab:digitizerParameters}).  Spatial resolution was modeled to improve with increasing interaction depth, corresponding to interactions closer to the photodetectors.

\begin{table*}[thb]
    \centering
    \caption{Prescribed detector-response models and parameters used by the external digitizer.  Spatial parameters were identical in the tangential, axial, and DOI directions within each specified resolution.  Energy \(E\) is expressed in MeV, and \(d\) is the interaction depth in mm.}
    \label{tab:digitizerParameters}
    \small
    \setlength{\tabcolsep}{5pt}
    \begin{tabularx}{\textwidth}{
        >{\raggedright\arraybackslash}p{0.17\textwidth}
        >{\raggedright\arraybackslash}X
        >{\raggedright\arraybackslash}p{0.43\textwidth}}
        \toprule
        Quantity & Response model & Parameters used\\
        \midrule

        Energy resolution
        &
        \(\displaystyle
        f_E(E)^2=
        \frac{a^2}{E}
        +\frac{b^2}{E^2}
        +c_E^2
        \)
        &
        Fractional FWHM;
        \(12\%\) at 511 keV;
        \(a=0.08088\), \(b=0\), \(c_E=0.04\);
        no depth dependence
        \\[3pt]

        Timing resolution
        &
        \(\displaystyle
        W_t(E)=W_{t,511}
        \sqrt{\frac{0.511}{E}}
        \)
        &
        \(W_{t,511}=141.4\) ps per photon detection;
        CTR \(=200\) ps FWHM at 511 keV
        \\[3pt]

        Spatial resolution
        &
        \(\displaystyle
        W_k(E,d)=
        \left[
        W_{k,511}^{(10)}
        +m_k(d-10~\mathrm{mm})
        \right]
        \sqrt{\frac{0.511}{E}}
        \)
        &
        \(k\in\{\mathrm{tan},\mathrm{ax},\mathrm{doi}\}\);
        \(W_{k,511}^{(10)}=1,2,3,4\) mm;
        reference depth \(d=10\) mm
        \\[3pt]

        Depth slope
        &
        \(\displaystyle
        m_k=-\frac{W_{k,511}^{(10)}}{80~\mathrm{mm}}
        \)
        &
        \(m_k=-0.0125,-0.025,-0.0375,-0.05\)
        mm FWHM/mm for the 1--4 mm conditions
        \\[3pt]

        Correlated response
        &
        \(\mathbf C=\mathbf D\boldsymbol{\Gamma}\mathbf D\)
        &
        \(\rho_{E,\tau}=-0.15\),
        \(\rho_{E,\mathrm{doi}}=-0.20\),
        \(\rho_{\tau,\mathrm{tan}}=0.10\),
        \(\rho_{\tau,\mathrm{doi}}=0.35\), and
        \(\rho_{\mathrm{tan},\mathrm{ax}}=0.05\);
        all other off-diagonal coefficients were zero
        \\[3pt]

        \bottomrule
    \end{tabularx}
\end{table*}

For each continuous measured data set, matched virtual-segmented measurements were generated with nominal tangential and axial spacings of \(2\), \(3\), and \(4\) mm.  The virtual detector coordinates were assigned at the face of the detector.  The continuous and segmented representations shared the same physical interactions, measured energies and times, and stochastic detector-response realization.

All projection models were evaluated using the same simulated detector responses for each resolution.  The models differed only in the measured information and response dependencies incorporated into their event kernels.

\subsubsection{Simulated objects and acquisitions}
\label{sec:simulatedObjects}

\begin{figure}[htb]
    \centering
    \includegraphics[width=0.45\linewidth]{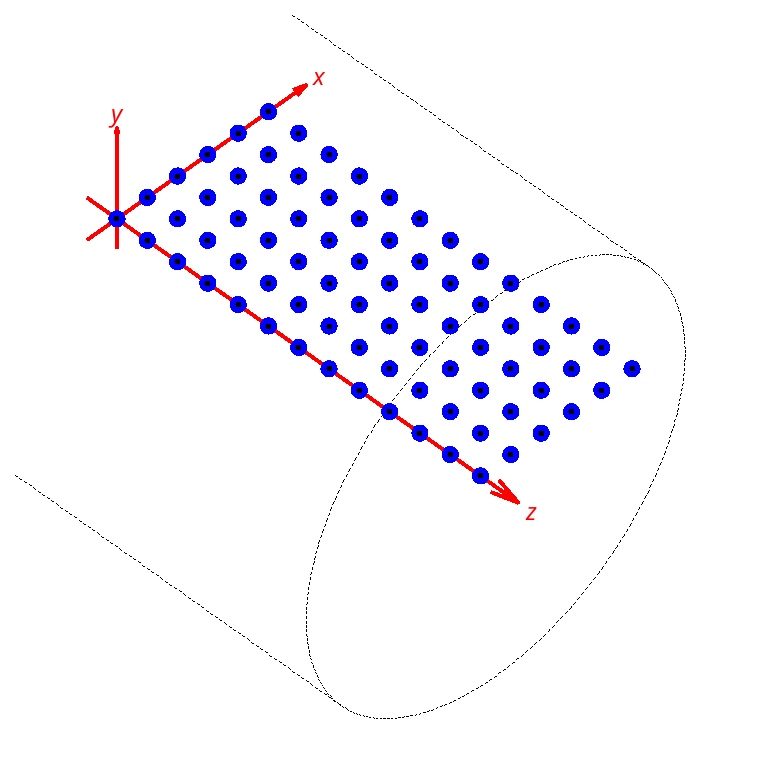}
    \includegraphics[width=0.525\linewidth]{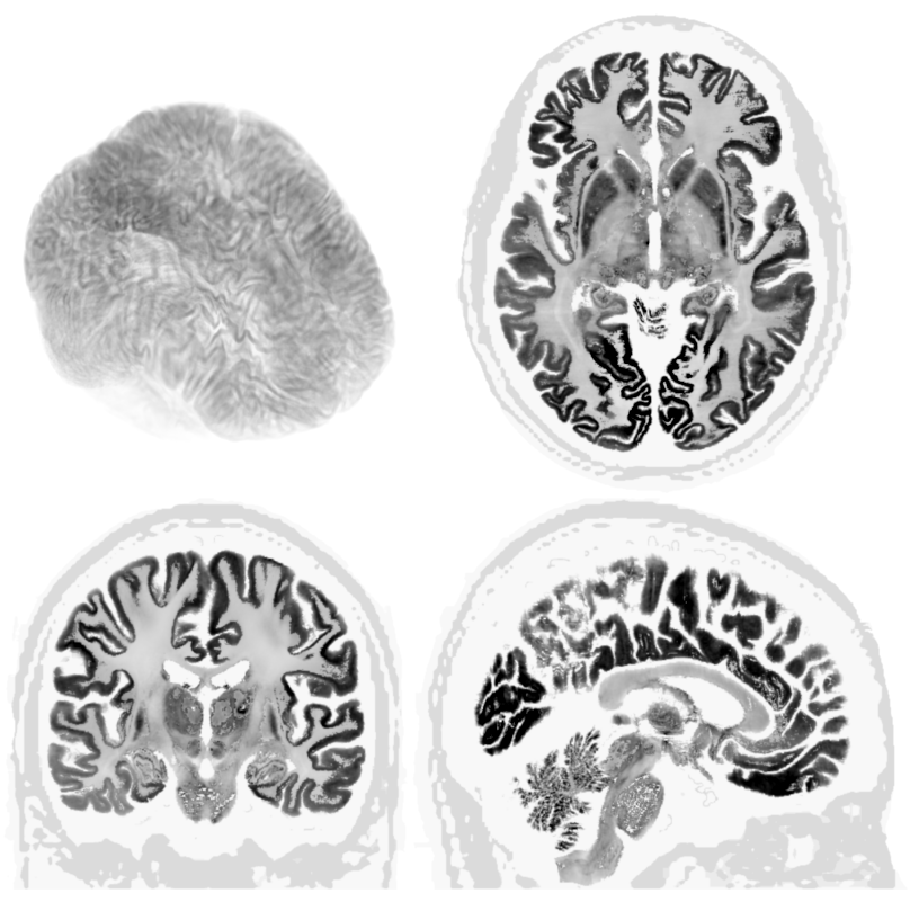}
    \caption{This figure shows the two phantoms we used to evaluate the SPLAT projection operator.  The left figure shows a perspective view of the $6\times13$ grid of points.  The right view shows the multi-view projection of an ultra high-resolution ($250\times250\times250~\mu$m voxels) of a T1-MR brain image.}
    \label{fig:gridMRI}
\end{figure}

Two types of \(^{18}\mathrm{F}\) data sets were simulated: a \(6\times13\) grid of 78 point-source locations for projection-model validation and a digital brain phantom for reconstruction evaluation (see Fig.~\ref{fig:gridMRI}).  The digital brain phantom was derived from a high resolution T1-weighted MRI data set ($250~\mu$m pixels)~\cite{lusebrink2017t1}.  To keep attenuation effects simple we used a uniform water mask based on the support of the MRI.  The acquisition parameters are summarized in Table~\ref{tab:simulatedAcquisitions}.

\begin{table}[htb]
    \centering
    \caption{Simulated objects and acquisition parameters.}
    \label{tab:simulatedAcquisitions}
    \small
    \begin{tabularx}{\columnwidth}{
        >{\raggedright\arraybackslash}p{0.16\columnwidth}
        >{\raggedright\arraybackslash}X
        >{\raggedright\arraybackslash}p{0.34\columnwidth}}
        \toprule
        Object & Geometry or activity distribution & Acquisition\\
        \midrule

        Point-source grid
        &
        78 (3.7~MBq each) point-source locations (\(6\times13\)), each at
        the center of a 4-mm-radius water sphere, with
        coordinates \((x,0,z)\),
        \(x=0{:}20{:}100\) mm and
        \(z=0{:}20{:}240\) mm
        &
        300 s;
        detector FWHM \(=1,2,3,4\) mm,
        with matched 2-, 3-, and 4-mm segmented representations
        \\[3pt]

        Digital brain phantom
        &
        $250~\mu$m isotropic T1-weighted MRI brain~\cite{lusebrink2017t1} with water attenuation; 37 MBq total activity,
        mean SUV \(=2\), maximum SUV \(=16\)
        & 300 s; nominal case: 3-mm detector FWHM with 3-mm segmentation;
        high-resolution case: 1-mm detector FWHM
        with 2-mm segmentation
        \\

        \bottomrule
    \end{tabularx}
\end{table}

\subsubsection{Projection models}
\label{sec:modelsCompared}

The projection models were designed to evaluate the contributions of different detector measurements to event localization.  The \modelSEG{p} models provided matched segmented comparators at spacings of \(p=2,3,4\) mm.  \modelBase used continuous detector coordinates with fixed, independent timing and spatial responses but without measured DOI.  \modelDOI added the measured DOI coordinate, \modelEnergy incorporated the energy-dependent and correlated detector response, and \modelDepth further included the depth dependence of the tangential, axial, and DOI resolutions.

For models without measured DOI, detection endpoints were assigned the reference depth \(d_{\mathrm{ref}}=10~\mathrm{mm}\).  The unresolved detector thickness \(T\) was approximated by a uniform depth distribution with variance \(\sigma_{\mathrm{uDOI}}^2=T^2/12\).  For segmented coordinates of spacing  \(\Delta_k\), the quantization variance was \(\sigma_{\mathrm{seg},k}^2=\Delta_k^2/12\).  Independent intrinsic localization and quantization variances were added.

The \modelSEG{p} models used the matched virtual-segmented measurements described above.  Measured DOI was discarded, and the segment variance was combined with the intrinsic detector-position variance.

\subsection{Projection-model validation and reconstruction evaluation}
\label{sec:projectionValidation}

\subsubsection{Point-source validation}

Projection-model performance was evaluated at 78 point-source locations for intrinsic detector resolutions of 1, 2, 3, and 4 mm FWHM.  To give each source equal weight, the number of accepted coincidences was limited to that available from the source with the fewest counts.  Events were divided equally into reference and test subsets.  All models used the same events, and their response parameters were specified independently of the test data.  Event-localization accuracy and agreement with the predicted Gaussian response were evaluated using the known GATE annihilation positions.

For event \(n\) and model \(M\), let \(\boldsymbol{\mu}_{Mn}\) and \(\boldsymbol{\Sigma}_{Mn}\) denote the mean and covariance of the Gaussian kernel \(\kappa_{\mathrm G,M}(m_n,\mathbf x)\) in~\eqref{eq:eventGaussianKernel}, where \(m_n\) is the descriptor of event \(n\).  The GATE annihilation position \(\mathbf x_n\) served as the reference position.  The localization residual, the normalized residual, and the squared Mahalanobis distance were defined as        
\begin{equation}
\begin{aligned}
    \mathbf r_{Mn}
      &:=
      \mathbf x_n-\boldsymbol{\mu}_{Mn},\qquad
    \boldsymbol{\Sigma}_{Mn}
      &=
      \mathbf L_{Mn}\mathbf L_{Mn}^{\mathsf T},\\
    \boldsymbol{\eta}_{Mn}
      &:=
      \mathbf L_{Mn}^{-1}\mathbf r_{Mn},\qquad
    d_{Mn}^{2}
      &:=
      \|\boldsymbol{\eta}_{Mn}\|_2^2 ,
\end{aligned}
\label{eq:validationResidualDefinitions}
\end{equation}
where \(\mathbf L_{Mn}\) is the Cholesky factor of \(\boldsymbol{\Sigma}_{Mn}\), so that \(\boldsymbol{\eta}_{Mn}\) expresses the residual in units of the kernel spread.  

The primary evaluation metric was the Gaussian negative log score (NLL) of the true
annihilation position,
\begin{equation}
\boxed{
    \ell_{Mn}:
      =
      -\log\kappa_{\mathrm G,M}(m_n,\mathbf x_n) 
      =
      \frac12
      \left[
        d_{Mn}^{2}
        +\log\det\boldsymbol{\Sigma}_{Mn}
        +3\log(2\pi)
      \right].
      }
\label{eq:validationNLL}
\end{equation}
Lower NLL indicates greater probability density assigned to the true annihilation position.  The score accounts for both localization error and the spatial extent of the predicted Gaussian response.  The use of logarithmic scores follows the proper-scoring-rule framework for evaluating probabilistic predictions~\cite{gneiting2007strictly}.

To express NLL in more familiar spatial units, we first consider an ideal case: an isotropic three-dimensional Gaussian kernel with per-axis standard deviation \(\sigma\), for which the normalized residual \(\boldsymbol{\eta}\) in~\eqref{eq:validationResidualDefinitions} has zero mean and unit variance in each direction.  Then \(d^2=\|\boldsymbol{\eta}\|_2^2\) has expectation \(\mathrm E[d^2]=3\), and \(\log\det\boldsymbol{\Sigma}=6\log\sigma\), so by~\eqref{eq:validationNLL} the expected NLL is
\[
    \mathrm E[\ell]
      =
      \frac12\bigl[3+6\log\sigma+3\log(2\pi)\bigr]
      =
      \frac32\log\!\left(2\pi e\,\sigma^{2}\right).
\]     
We set this expected NLL equal to the mean NLL of model \(M\),
\[
    \overline{\ell}_M
      =
      \frac32\log\!\left(2\pi e\,\sigma^{2}\right),
\]                                                          
where \(\overline{\ell}_M\) is the mean NLL over the equally represented source locations, with spatial coordinates expressed in millimeters.  Solving for \(\sigma\) gives
\begin{equation}
\boxed{
    \sigma_{\mathrm{NLL},M}
      :=
      \frac{\exp\!\left(\overline{\ell}_M/3\right)}{\sqrt{2\pi e}}.
}                                                          
\label{eq:validationSigmaNLL}
\end{equation}
Thus \(\sigma_{\mathrm{NLL},M}\) is the standard deviation of the ideal Gaussian that has the same expected NLL as model \(M\).  It provides an equivalent localization scale while preserving the NLL ranking of the projection models.  

\subsubsection{Localization errors and Gaussian-reference diagnostics}

Physical localization errors were evaluated along and perpendicular to the measured LOR.  The residual components were defined as
\begin{equation}
    r_{\parallel,Mn}
      :=
      \mathbf e_n^{\mathsf T}\mathbf r_{Mn},
    \qquad
    \mathbf r_{\perp,Mn}
      :=
      \mathbf r_{Mn}
      -
      r_{\parallel,Mn}\mathbf e_n .
\label{eq:lorResidualComponents}
\end{equation}
The localization errors were summarized by
\begin{equation}
\begin{aligned}
    \operatorname{MAE}_{\parallel}
      &:=
      \mathrm E[|r_{\parallel}|],
    &
    \sigma_{\parallel}
      &:=
      \operatorname{SD}(r_{\parallel}),\\
    \mu_{\perp}
      &:=
      \mathrm E[\|\mathbf r_{\perp}\|_2],
    &
    \operatorname{RMS}_{\perp}
      &:=
      \sqrt{\mathrm E[\|\mathbf r_{\perp}\|_2^2]},
\end{aligned}
\label{eq:lorResidualMetrics}
\end{equation}
where \(\mathrm E[\cdot]\) and \(\operatorname{SD}(\cdot)\) denote the mean and the standard deviation over the analyzed events.  The standard deviations of the two transverse residual components, \(\sigma_{\perp t}\) and \(\sigma_{\perp z}\), were also recorded.

Agreement with the predicted Gaussian response was evaluated using the covariance-normalized residuals \(\boldsymbol{\eta}_{Mn}\).  For an ideal three-dimensional Gaussian response, these residuals have zero mean and identity covariance, while their squared Mahalanobis distances follow 
\[
    d^2\sim\chi_3^2.
\]
We evaluated the mean and covariance of the normalized residuals, the mean and standard deviation of \(d^2\), and the empirical coverage of the nominal 95\% and 99\% Gaussian probability regions.  These measures were used to assess agreement with the Gaussian reference distribution, while model comparisons were based on NLL.

\subsubsection{Model comparisons}

Model differences were evaluated using paired measurements from the same events at each source location.  To avoid weighting locations according to their detection sensitivity, the number of evaluated events was equalized across the 78 sources.

For a metric \(Y\), let \(Y_{M,q}\) denote its value for model \(M\) at source \(q\), and define the paired difference \(\Delta_{A-B,q}=Y_{A,q}-Y_{B,q}\).  The mean difference and its source-level standard error were 
\begin{equation}
    \overline{\Delta}_{A-B}
      =
      \frac{1}{78}
      \sum_{q=1}^{78}\Delta_{A-B,q},
    \qquad
    \operatorname{SE}_{\mathrm{src}}
      =
      \frac{s_{\Delta}}{\sqrt{78}},
\label{eq:sourceBalancedContrast}
\end{equation}
where \(s_{\Delta}\) is the sample standard deviation of the 78 source-level differences (using 77 degrees of freedom).  Thus, the standard error is based on variation across source locations rather than individual events.

The incremental comparisons were \modelBase\(\rightarrow\)\modelDOI, \modelDOI\(\rightarrow\)\modelEnergy, and \modelEnergy\(\rightarrow\)\modelDepth.  Each \modelSEG{p} model was also compared with \modelBase and \modelDepth.  Comparisons were performed separately for each intrinsic detector resolution.

\subsubsection{Information-equivalent count gain}

Improved event localization can increase the spatial information provided by each accepted coincidence.  We characterize the spatial concentration of the Gaussian event response by
\begin{equation}
    q_M(m)
      :=
      \int_{\mathbb R^3}
      \kappa_{\mathrm G}^2(m,\mathbf x)\,d\mathbf x
      =
      \frac{1}{
        (4\pi)^{3/2}
        \sqrt{\det\boldsymbol{\Sigma}_M(m)}
      }.
\label{eq:count_quality}
\end{equation}
A more concentrated response has a larger \(q_M\), reflecting the greater spatial information potentially provided by that event.

The connection to information can be demonstrated by considering a weak point source at location \(\mathbf x_0\), superimposed on a background activity distribution \(f_0\).  Following the list-mode Fisher-information work of \citet{barrett1995objective, barrett1997list, parra1998list}, the Poisson Fisher information for the source amplitude is
\begin{equation}
    \mathcal I_M(\mathbf x_0;f_0)
      =
      \int_{\mathcal M}
      \frac{K_M^2(m,\mathbf x_0)}
           {g_M(m;f_0)}\,dm  \, .
\label{eq:weakPointSourceInformation}
\end{equation}
For a uniform background \(f_0=\rho\)negligible additive contamination, and negligible kernel truncation at the image-domain boundary, \(g_M(m;f_0)\simeq\rho w_M(m)\).  Averaging the Fisher information over source positions gives
\begin{equation}
    \frac{1}{|\Omega|}
      \int_{\Omega}\mathcal I_M(\mathbf x_0;f_0)\,d\mathbf x_0
      \simeq
      \frac{\overline N_M}{\rho^2|\Omega|}
      \langle q_M\rangle_{w_M},\quad \text{with} \quad
    \langle q_M\rangle_{w_M}
      :=
      \frac{\int_{\mathcal M}w_M(m)q_M(m)\,dm}
           {\int_{\mathcal M}w_M(m)\,dm},
\label{eq:meanWeakPointSourceInformation}
\end{equation}
where \(\overline N_M=\rho\int_{\mathcal M}w_M(m)\,dm\) is the expected accepted count.  Thus, under these assumptions, spatially averaged weak-point-source information is proportional to the accepted count multiplied by the weighted mean response concentration.  This motivates using \(q_M\) to compare event-localization information.

To approximately account for positron range and finite voxel dimensions, we introduce the effective covariance
\begin{equation}
    \boldsymbol{\Sigma}_{\mathrm{eff},M}(m)
      =
      \boldsymbol{\Sigma}_M(m)
      +
      \boldsymbol{\Sigma}_{\mathrm{pr}}
      +
      \boldsymbol{\Sigma}_{\mathrm{vox}},
\label{eq:effectiveEventCovariance}
\end{equation}
where \(\boldsymbol{\Sigma}_{\mathrm{pr}} =\sigma_{\mathrm{pr}}^2\mathbf I_3\) represents positron-range uncertainty (\(\sigma_{\mathrm{pr}}=0.29\) mm for \(^{18}\mathrm F\) in water~\cite{carter2020impact}) and \(\boldsymbol{\Sigma}_{\mathrm{vox}} = \frac{1}{12}  \operatorname{diag} (\Delta_x^2,\Delta_y^2,\Delta_z^2)\) represents the covariance of a uniform distribution over the image voxel.  The corresponding Gaussian-equivalent response concentration is
\begin{equation}
    q_{\mathrm{eff},M}(m)
      =
      \frac{1}{
        (4\pi)^{3/2}
        \sqrt{
          \det\boldsymbol{\Sigma}_{\mathrm{eff},M}(m)
        }
      }.
\label{eq:voxelAdjustedCountQuality}
\end{equation}
These covariance additions approximate the effects of positron range and image discretization on spatial concentration without modifying the SPLAT projection kernel itself.

For configurations \(A\) and \(B\), we define the approximate information-equivalent count gain as
\begin{equation}
\boxed{
    G_{B:A}
      :=
      \frac{\epsilon_{\mathrm{NEC},B}}
           {\epsilon_{\mathrm{NEC},A}}
      \frac{\overline q_{\mathrm{eff},B}}
           {\overline q_{\mathrm{eff},A}}.
}
\label{eq:count_equivalent_gain}
\end{equation}
Here, \(\epsilon_{\mathrm{NEC},M}\) is the noise-equivalent count efficiency, defined as noise-equivalent counts divided by the number of emitted annihilation pairs.  Noise-equivalent counts account approximately for the statistical effects of scattered and random coincidences \cite{Strother_1990}.  The quantity \(\overline q_{\mathrm{eff},M}\) is calculated by averaging the response concentration within each source location, with equal weighting across source locations.

The first factor describes relative counting efficiency, while the second describes relative spatial concentration.  For the matched true-coincidence data used in this study, the efficiency ratio is unity.  The resulting gain provides an approximate measure of information-equivalent counts; its relationship to reconstructed-image quality depends on the background distribution, feature characteristics, and imaging task.

\subsubsection{Reconstruction evaluation}
\label{sec:projectionReconstructionDemonstrations}

Representative Gaussian event kernels were visualized to illustrate the effects of detector resolution and the projection model on their spatial extent, position, and orientation.

The 78-point-source grid was reconstructed to evaluate localization accuracy in image space.  For each reconstructed source, the mean position and standard deviation were estimated along the three Cartesian axes within a local region of the reconstructed image.  Localization bias was defined as the difference between the reconstructed and true source positions.  The combined root-mean-square error was calculated as
\begin{equation}
    \mathrm{RMSE}
      =
      \sqrt{
        \sum_{i\in\{x,y,z\}}
        \left(
          \Delta\mu_i^2+\sigma_i^2
        \right)
      },
\label{eq:reconstructedPointRMSE}
\end{equation}
where \(\Delta\mu_i\) and \(\sigma_i\) denote the reconstructed source-position bias and spatial spread, respectively.  Reconstructions were performed without regularization and evaluated after 5 iterations with 5 subsets.

The digital brain phantom was reconstructed to evaluate the effects of the projection model on image quality.  Comparisons used \modelDepth and the corresponding \modelBase and segmented models for 1-mm intrinsic detector resolution.  Within each comparison, the same accepted coincidences, measured energies and times, image grid, initialization, and reconstruction parameters were used.  Reconstructions employed the method described in section~\ref{sec:numericalReconstruction}.  Reconstructions were performed with higher-order total-variation (HOTV) regularization and evaluated after 25 iterations with 25 subsets.  The first- and second-order penalty weights were individually optimized by golden search.  When combined their values were halved.

\section{Results}
\label{sec:results}

The results are presented in five parts.  First, the event-specific projection models are evaluated against known annihilation positions from GATE Monte Carlo simulations, providing a direct assessment of localization accuracy and agreement with the predicted Gaussian responses.  Visualizations of these responses are shown.  We then examine the information-equivalent count gain associated with the different detector-response models.  Next, point-source and brain-phantom reconstructions are used to evaluate how differences in the projection models affect reconstructed-image accuracy and quality.  Finally, we report the computational performance of the GPU forward- and back-projection operators.

\subsection{Point-grid projection-model validation}
\label{sec:resultsPointGrid}

The projection models were evaluated by comparing their predicted event-localization distributions with the known annihilation positions from GATE using the point grid phantom.  We used annihilation rather than radioactive-decay positions to isolate the accuracy of the projection models from positron-range effects, which are treated separately.

\paragraph{Predictive localization.}
Approximately \(5.53\times10^{6}\) events distributed equally across the 78 point-source locations were evaluated for each model and intrinsic detector resolution.  The Gaussian negative log score (NLL) was lowest for \modelDepth at every resolution.  Relative to \modelDepth, the segmented models showed increasing NLL penalties with larger segment spacing, with the largest differences at the highest intrinsic detector resolution (Table~\ref{tab:nllContrasts}, part A).  Among the continuous models, measured DOI provided the largest improvement, with smaller additional gains from energy-dependent correlations and depth-dependent detector resolution (part B).

\begin{table}[htb]
    \centering
    \caption{Paired differences in Gaussian negative log score (NLL) per event, expressed as mean \(\pm\) source-level standard error across 78 source locations.  Each difference is calculated as the NLL of the first model minus that of the second model. Arrows in part B indicate the direction of the model comparison.  Positive values therefore favor the second model.}
    \label{tab:nllContrasts}
    \small
    \setlength{\tabcolsep}{4.5pt}
    \renewcommand{\arraystretch}{0.95}
    \begin{tabularx}{0.94\linewidth}{@{} C{3cm}|*{3}{Y} @{} }
        \toprule
        \multicolumn{4}{c}{\textbf{A.  Segmented models relative to \modelDepth}}\\
        \midrule
        Intrinsic FWHM
        & \modelSEG{2}\(-\)\modelDepth
        & \modelSEG{3}\(-\)\modelDepth
        & \modelSEG{4}\(-\)\modelDepth\\
        \midrule
        1 mm & \(2.062\pm0.055\) & \(2.283\pm0.047\) & \(2.499\pm0.041\)\\
        2 mm & \(1.411\pm0.046\) & \(1.563\pm0.041\) & \(1.732\pm0.037\)\\
        3 mm & \(1.016\pm0.037\) & \(1.118\pm0.034\) & \(1.242\pm0.031\)\\
        4 mm & \(0.770\pm0.030\) & \(0.841\pm0.028\) & \(0.933\pm0.026\)\\
        \midrule
        \multicolumn{4}{c}{\textbf{B.  Incremental improvements in continuous models}}\\
        \midrule
        Intrinsic FWHM
        & \modelBase\(\rightarrow\)\modelDOI
        & \modelDOI\(\rightarrow\)\modelEnergy
        & \modelEnergy\(\rightarrow\)\modelDepth\\
        \midrule
        1 mm & \(1.759\pm0.067\) & \(0.0137\pm0.0009\) & \(0.0169\pm0.0017\)\\
        2 mm & \(1.228\pm0.051\) & \(0.0206\pm0.0005\) & \(0.0092\pm0.0005\)\\
        3 mm & \(0.894\pm0.039\) & \(0.0216\pm0.0005\) & \(0.0064\pm0.0002\)\\
        4 mm & \(0.681\pm0.031\) & \(0.0214\pm0.0005\) & \(0.0047\pm0.0001\)\\
        \bottomrule
    \end{tabularx}
\end{table}

The corresponding NLL-equivalent localization scales, \(\sigma_{\mathrm{NLL}}\), are summarized in Table~\ref{tab:sigmaNLL}.  At 1-mm intrinsic detector resolution, \(\sigma_{\mathrm{NLL}}\) was 1.36 mm for \modelDepth, compared with 2.46 mm for \modelBase and 2.70, 2.91, and 3.12 mm for \modelSEG{2}, \modelSEG{3}, and \modelSEG{4}, respectively.

\begin{table*}[htb]
    \centering
    \caption{NLL-equivalent localization \(\sigma_{\mathrm{NLL}}\) (mm).  Smaller values correspond to better predictive localization.}
    \label{tab:sigmaNLL}
    \small
    \setlength{\tabcolsep}{3.5pt}
    \renewcommand{\arraystretch}{0.95}
    \begin{tabular}{@{} C{2.1cm}|*{7}{C{1.5cm}} @{} }
        \toprule
        Intrinsic FWHM
        & \modelSEG{2}
        & \modelSEG{3}
        & \modelSEG{4}
        & \modelBase
        & \modelDOI
        & \modelEnergy
        & \modelDepth\\
        \midrule
        1 mm & 2.70 & 2.91 & 3.12 & 2.46 & 1.37 & 1.37 & 1.36\\
        2 mm & 2.97 & 3.12 & 3.31 & 2.82 & 1.87 & 1.86 & 1.86\\
        3 mm & 3.29 & 3.40 & 3.55 & 3.19 & 2.37 & 2.35 & 2.34\\
        4 mm & 3.62 & 3.70 & 3.82 & 3.54 & 2.82 & 2.80 & 2.80\\
        \bottomrule
    \end{tabular}
\end{table*}

\paragraph{Gaussian-reference diagnostics.}
The normalized residuals for \modelDepth showed little mean bias, with \(\|\overline{\boldsymbol\eta}\|_2\) ranging from 0.005 to 0.017 (Table~\ref{tab:depthCalibration}).  The mean squared Mahalanobis distance approached the ideal Gaussian value of three as intrinsic detector resolution became coarser.  Empirical 95\% and 99\% coverage remained close to their nominal values.  However, the standard deviation of \(d^2\) exceeded the Gaussian reference, particularly at 1-mm intrinsic detector resolution, indicating departures from Gaussian behavior in the distribution tails (see Discussion section).

\begin{table}[htb]
    \centering
    \caption{Gaussian-reference diagnostics for \modelDepth events.  For an ideal three-dimensional Gaussian response, \(\|\overline{\boldsymbol\eta}\|_2=0\), \(\mathrm E[d^2]=3\), \(\operatorname{SD}(d^2)=\sqrt 6\approx2.45\), and coverage equals its nominal value.}
    \label{tab:depthCalibration}
    \small
    \setlength{\tabcolsep}{5pt}
    \begin{tabular}{@{} C{2.5cm}|*{5}{C{1.25cm}} @{} }
        \toprule
        Intrinsic FWHM
        & \(\|\overline{\boldsymbol\eta}\|_2\)
        & Mean \(d^2\)
        & SD \(d^2\)
        & 95\% coverage
        & 99\% coverage\\
        \midrule
        1 mm & 0.005 & 3.41 & 26.0  & 0.950 & 0.987\\
        2 mm & 0.008 & 3.11 &  8.63 & 0.951 & 0.989\\
        3 mm & 0.012 & 3.04 &  4.61 & 0.951 & 0.989\\
        4 mm & 0.017 & 3.01 &  3.32 & 0.951 & 0.989\\
        \bottomrule
    \end{tabular}
\end{table}

\paragraph{Single-event localization errors.}
Table~\ref{tab:lorLocalization} summarizes localization errors parallel and perpendicular to each event's measured LOR.  The longitudinal errors were similar for the segmented and \modelDepth models, consistent with their common TOF measurements.  The differences were substantially larger in the transverse plane.  At 1-mm intrinsic resolution, the transverse RMS error decreased from 2.54 mm for \modelSEG{2} to 0.63 mm for \modelDepth.  At 4-mm intrinsic resolution, the corresponding values were 3.22 and 1.90 mm. The largest improvements therefore occurred at the highest intrinsic detector resolution.

\begin{table}[htb]
    \centering
    \caption{Single-event localization errors (mm) in the event-specific LOR frame, with matched segmented comparators.  Values are equal-source summaries across 78 positions.}
    \label{tab:lorLocalization}
    \small
    \setlength{\tabcolsep}{3.4pt}
    \renewcommand{\arraystretch}{0.95}
    \begin{tabular}{@{}llcccccc@{}}
        \toprule
        & & \multicolumn{2}{c}{Parallel to LOR}
            & \multicolumn{4}{c}{Perpendicular to LOR}\\
        \cmidrule(lr){3-4}\cmidrule(lr){5-8}
        FWHM & Model
        & \(\operatorname{MAE}_{\parallel}\)
        & \(\sigma_{\parallel}\)
        & \(\mu_{\perp}\)
        & \(\sigma_{\perp t}\)
        & \(\sigma_{\perp z}\)
        & \(\operatorname{RMS}_{\perp}\)\\
        \midrule
        1 mm & \modelSEG{2} & 10.5 & 13.1 & 2.05 & 1.70 & 1.62 & 2.54\\
        & \modelDepth  &  10.0 & 12.5 & 0.50 & 0.45 & 0.44 & 0.63\\[3pt]
        2 mm & \modelSEG{2} & 10.5 & 13.1 & 2.19 & 1.79 & 1.71 & 2.65\\
        & \modelDepth  &  9.93 & 12.5 & 0.87 & 0.72 & 0.72 & 1.02\\[3pt]
        3 mm & \modelSEG{3} & 10.5 & 13.1 & 2.47 & 1.99 & 1.89 & 2.90\\
        & \modelDepth  &  9.89 & 12.4 & 1.27 & 1.03 & 1.03 & 1.45\\[3pt]
        4 mm & \modelSEG{4} & 10.5 & 13.1 & 2.79 & 2.24 & 2.12 & 3.22\\
        & \modelDepth  &  9.86 & 12.4 & 1.67 & 1.35 & 1.34 & 1.90\\
        \bottomrule
    \end{tabular}
\end{table}

\subsection{Event-specific Gaussian response visualization}
\label{sec:resultsKernelVisualization}

\begin{figure}[htb]
    \centering
    \includegraphics[width=\linewidth]{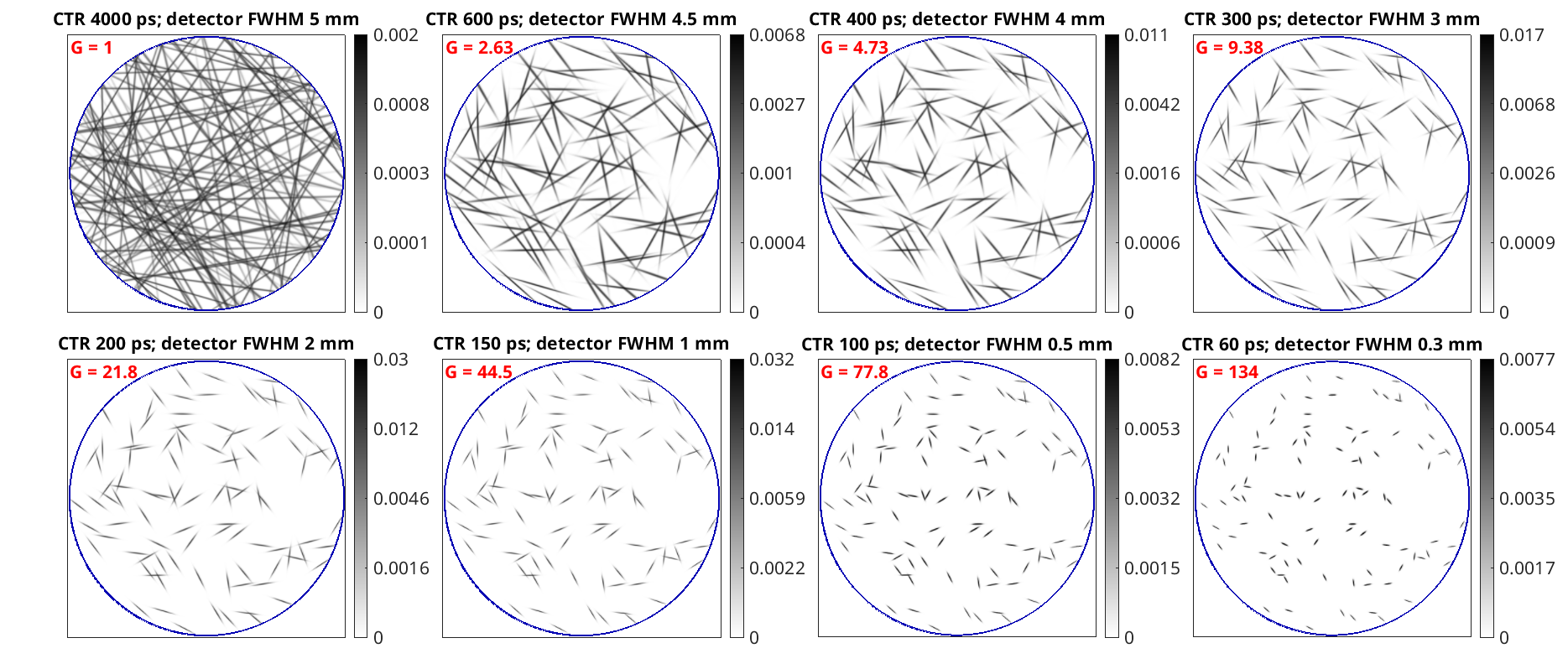}
    \caption{Gaussian splats for 100 photoelectric coincidence events in an 800-mm-diameter (whole body PET), 20-mm-thick LSO ring, displayed over a 700-mm field of view.  Panels show progressively improved timing and spatial resolution; their annotations give relative  information-equivalent count gain estimated for 1-mm\(^3\) voxels and \(^{18}\mathrm F\) positron range.}
    \label{fig:splatCompare}
\end{figure}

\begin{figure}[htb]
    \centering
    \includegraphics[width=\linewidth]{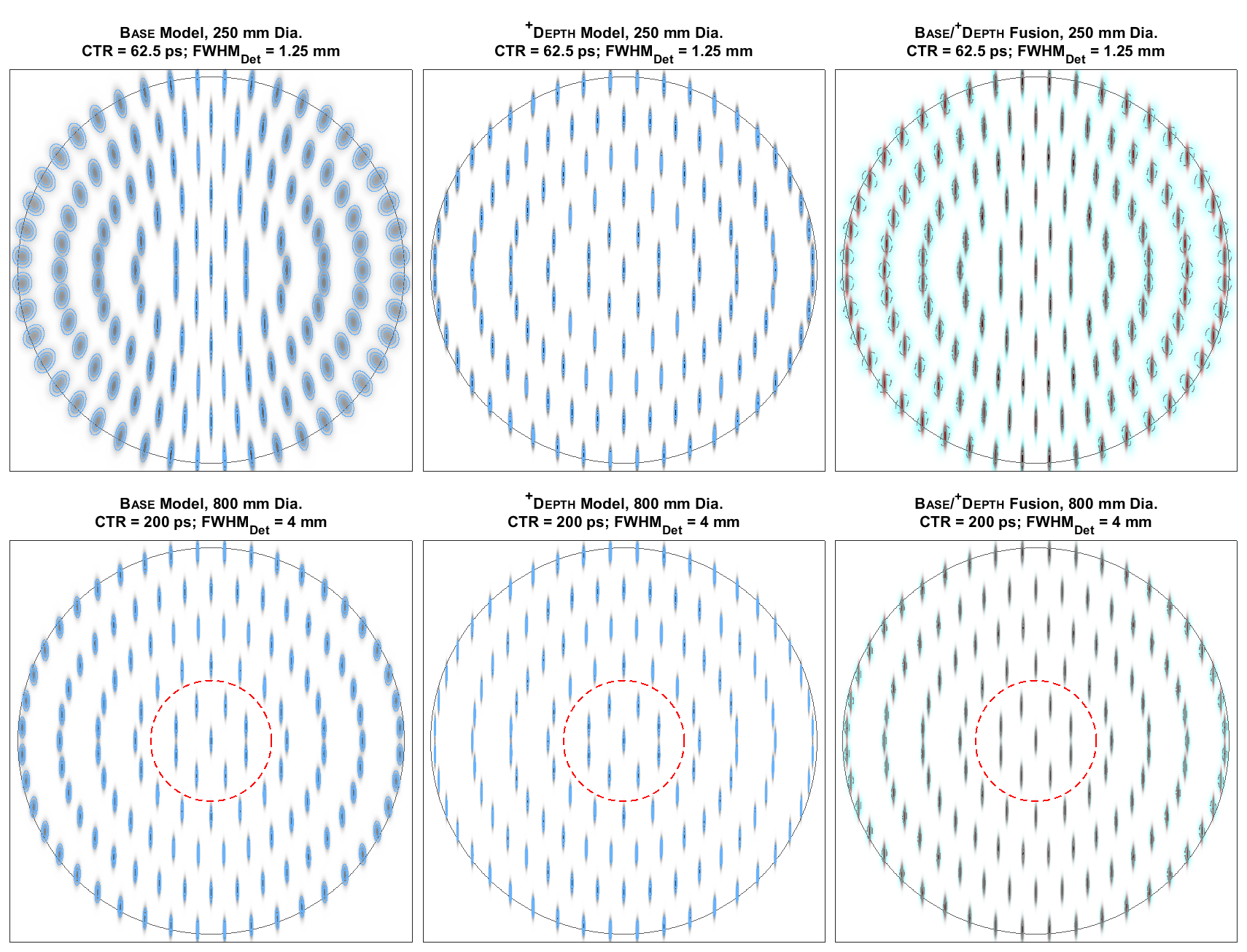}
    \caption{\modelBase and \modelDepth event responses for vertical LORs in 250-mm (top) and 800-mm (bottom) diameter scanners. The respective timing resolutions of 62.5 and 200 ps FWHM, and detector spatial resolutions of 1.25 and 4 mm FWHM, were chosen to illustrate differences between the model responses.  Overlays compare the responses, with contours at the indicated relative levels.  We use a dashed red circle to illustrate the size difference of the upper and lower FOV, 250 and 800 mm, respectively.}
    \label{fig:splatPSFCompare}
\end{figure}

Figures~\ref{fig:splatCompare} and~\ref{fig:splatPSFCompare} illustrate the spatial characteristics of the event-specific Gaussian responses using representative PET scanner geometries.

Figure~\ref{fig:splatCompare} shows the effects of timing and detector spatial resolution on the responses of the same 100 coincidences in an 800-mm-diameter, whole-body-scale detector ring.  The panel annotations give the corresponding information-equivalent count gains, calculated for 1-mm isotropic voxels and an \(^{18}\mathrm F\) positron-range model.

Figure~\ref{fig:splatPSFCompare} compares \modelBase and \modelDepth responses in 250- and 800-mm-diameter scanner geometries, illustrating the effects of DOI information on spatial localization.

\subsection{Information-equivalent count gain}
\label{sec:resultsCountGain}

Information-equivalent count gain was calculated from the event-specific covariance matrices for the 78-point-source Monte Carlo simulations.  The calculations included Gaussian-equivalent corrections for \(^{18}\mathrm F\) positron range in water and 1-mm isotropic voxels.  Because each comparison used the same accepted coincidences, the count-efficiency ratio was unity.

Table~\ref{tab:grid-count-gain} summarizes the source-balanced information-equivalent gains across the four intrinsic detector resolutions.  Comparisons include \modelDepth relative to the \modelBase and matched segmented references.

\begin{table}[htb]
\centering
\caption{Information-equivalent count gains for the projection models evaluated using the 78-point-source Monte Carlo simulations.  Each entry \(G_{B:A}\) represents the gain of model \(B\) relative to reference model \(A\), calculated from the source-balanced mean effective Gaussian response concentrations via \eqref{eq:count_equivalent_gain}; Fig.~\ref{fig:Point-Grid-phantom} provides spatial context for these averages.  Values greater than unity indicate greater spatial response concentration for model \(B\).  \modelBase and \modelDepth are both continuous-coordinate models; \modelDepth additionally incorporates measured DOI, energy-dependent correlations, and depth-dependent detector resolution.  The calculations assume \(^{18}\mathrm F\) positron range in water and 1-mm isotropic voxels.  All comparisons use matched accepted coincidences, so the count-efficiency ratio is unity.}
\label{tab:grid-count-gain}
\small
\setlength{\tabcolsep}{3pt}
\renewcommand{\arraystretch}{0.9}
\begin{tabular}{@{}ccccc@{}}
\toprule
Intrinsic & & $G_{\modelDepth:\modelBase}$ & 
\multicolumn{2}{c@{}}{$G_{\modelDepth:\modelSEG{p}}$} \\
\cmidrule(lr){3-3}\cmidrule(l){4-5}
 FWHM & SEG ref. & Adj. & Raw & Adj. \\
\midrule
1 mm & \modelSEG{2} & 3.65 & 7.79 & 4.76 \\
2 mm & \modelSEG{2} & 2.68 & 3.70 & 3.15 \\
3 mm & \modelSEG{3} & 2.11 & 2.83 & 2.62 \\
4 mm & \modelSEG{4} & 1.78 & 2.39 & 2.28 \\
\bottomrule
\end{tabular}
\end{table}

\subsection{Reconstruction results}
\label{sec:resultsRecon}

Two phantoms were used to evaluate image reconstruction: a 78-point-source grid and a digital brain phantom derived from 250-$\mu$m-resolution T1 MRI.  The point-grid simulation used 3.7 MBq of $^{18}$F per source over 300 s, producing approximately 105 million accepted PE coincidences.  This phantom was reconstructed on a $500\times500\times1000$ matrix using 5 subsets in 5 iterations, without post-filtering or regularization.  The brain phantom, with 37 MBq of $^{18}$F, was simulated for 300 s and produced approximately 22.4 million accepted PE coincidences.  It was reconstructed on a $250\times250\times500$ grid using 25 iterations with 25 subsets and higher-order total variation (HOTV) regularization as described in~\cite{ross2017relaxed,li2015effective,lin2019krasnoselskii}.

\subsubsection{Point-grid reconstruction}
\label{sec:resultsPointGridRecon}

\begin{figure}[htb]
    \centering
    \includegraphics[width=\linewidth]{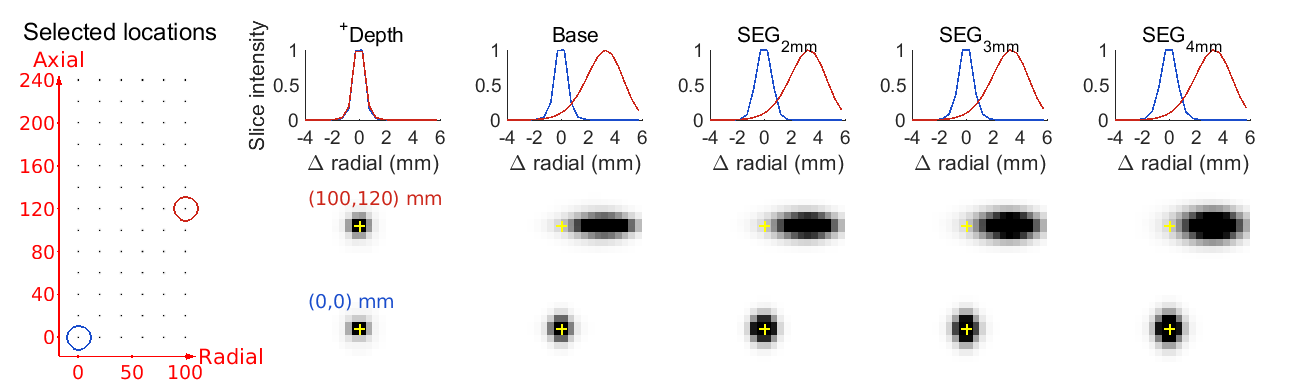}
    \caption{Point-grid reconstructions at 1-mm intrinsic detector resolution, comparing \modelDepth, \modelBase, and \modelSEG{2,3,4}.  Coronal enlargements and profiles at two selected locations illustrate spatial spread and localization bias; corresponding metrics are reported in Table~\ref{tab:grid-pkma-fwhm-combined}.}
    \label{fig:Point-Grid-phantom}
\end{figure}

Reconstructed source localization was evaluated using the 78-point-source grid, comparing \modelDepth with \modelBase and the three matched segmented models.  All available accepted coincidences were used, with identical reconstruction settings for each model at a given intrinsic detector resolution.  Reconstructed source-profile FWHM, position bias, and combined RMSE are summarized in Table~\ref{tab:grid-pkma-fwhm-combined} for representative sources at the center and near the radial limits of the FOV.  Representative reconstructions at 1-mm intrinsic detector resolution are shown in Fig.~\ref{fig:Point-Grid-phantom}.

At the center of the FOV, all models showed negligible localization bias, although \modelDepth generally produced narrower reconstructed source profiles.  At the peripheral source \((100,0,0)\) mm, the differences were substantially larger.  Models without measured DOI showed positive radial biases of 3.09--3.19 mm across the four intrinsic detector resolutions, compared with 0.014--0.174 mm for \modelDepth.  The corresponding radial FWHM values were 4.15--4.41 mm without measured DOI and 1.12--1.74 mm for \modelDepth.

These differences were reflected in the combined RMSE.  At 1-mm intrinsic detector resolution, peripheral-source RMSE decreased from 3.52 mm for \modelSEG{2} to 0.99 mm for \modelDepth, a reduction of approximately 72\%.  At the central source, the corresponding values were 1.02 and 0.94 mm.  The peripheral differences persisted across all four intrinsic detector resolutions and were consistent with the expected effects of DOI-related parallax in the cylindrical detector geometry.

\begin{table}[htb]\centering\small
\caption{Reconstructed point-source profile FWHM in the radial (\(r\)), tangential (\(t\)), and axial (\(z\)) directions, radial centroid bias \(\Delta\mu_r\), and combined RMSE for sources at the scanner center and 100 mm radial displacement. All values are in mm.}
\label{tab:grid-pkma-fwhm-combined}
\setlength{\tabcolsep}{3pt}\renewcommand{\arraystretch}{1}
\renewcommand{\arraystretch}{0.95}
\begin{tabular}{@{}>{\raggedright\arraybackslash}p{\modelwidth}*{5}{>{\centering\arraybackslash}p{\numberwidth}}@{}p{\sourcegap}@{}*{5}{>{\centering\arraybackslash}p{\numberwidth}}@{}}
\toprule
\multicolumn{12}{@{}l}{\textbf{Intrinsic FWHM: 1 mm}} \\[2pt]
 & \multicolumn{5}{c}{Source at $(0,0,0)$ mm} & {} & \multicolumn{5}{c}{Source at $(100,0,0)$ mm} \\
\cmidrule(lr){2-6}\cmidrule(lr){8-12}
Model & {\scriptsize FWHM$_r$} & {\scriptsize FWHM$_t$} & {\scriptsize FWHM$_z$} & $\Delta\mu_r$ & RMSE & {} & {\scriptsize FWHM$_r$} & {\scriptsize FWHM$_t$} & {\scriptsize FWHM$_z$} & $\Delta\mu_r$ & RMSE \\\midrule
\modelDepth & 1.10 & 1.10 & 1.06 & 0.00 & 0.94 & {} & 1.12 & 1.11 & 1.12 & 0.01 & 0.99 \\
\modelBase & 1.17 & 1.17 & 1.22 & 0.00 & 0.98 & {} & 4.15 & 1.29 & 1.27 & 3.09 & 3.49 \\
\modelSEG{2} & 1.42 & 1.42 & 1.37 & -0.00 & 1.02 & {} & 4.16 & 1.58 & 1.45 & 3.12 & 3.52 \\
\modelSEG{3} & 1.29 & 1.29 & 1.46 & 0.00 & 1.08 & {} & 4.20 & 1.85 & 1.66 & 3.14 & 3.57 \\
\modelSEG{4} & 1.43 & 1.44 & 1.58 & 0.00 & 1.16 & {} & 4.27 & 2.12 & 1.85 & 3.17 & 3.64 \\
\midrule
\multicolumn{12}{@{}l}{\textbf{Intrinsic FWHM: 2 mm}} \\[2pt]
 & \multicolumn{5}{c}{Source at $(0,0,0)$ mm} & {} & \multicolumn{5}{c}{Source at $(100,0,0)$ mm} \\
\cmidrule(lr){2-6}\cmidrule(lr){8-12}
Model & {\scriptsize FWHM$_r$} & {\scriptsize FWHM$_t$} & {\scriptsize FWHM$_z$} & $\Delta\mu_r$ & RMSE & {} & {\scriptsize FWHM$_r$} & {\scriptsize FWHM$_t$} & {\scriptsize FWHM$_z$} & $\Delta\mu_r$ & RMSE \\\midrule
\modelDepth & 1.28 & 1.28 & 1.24 & -0.00 & 0.96 & {} & 1.26 & 1.38 & 1.31 & 0.05 & 1.01 \\
\modelBase & 1.39 & 1.39 & 1.49 & 0.00 & 1.09 & {} & 4.19 & 1.64 & 1.51 & 3.12 & 3.55 \\
\modelSEG{2} & 1.50 & 1.51 & 1.63 & -0.00 & 1.14 & {} & 4.20 & 1.84 & 1.66 & 3.14 & 3.58 \\
\modelSEG{3} & 1.55 & 1.55 & 1.76 & 0.00 & 1.21 & {} & 4.24 & 2.03 & 1.80 & 3.15 & 3.62 \\
\modelSEG{4} & 1.65 & 1.65 & 1.82 & -0.00 & 1.29 & {} & 4.29 & 2.25 & 1.97 & 3.17 & 3.68 \\
\midrule
\multicolumn{12}{@{}l}{\textbf{Intrinsic FWHM: 3 mm}} \\[2pt]
 & \multicolumn{5}{c}{Source at $(0,0,0)$ mm} & {} & \multicolumn{5}{c}{Source at $(100,0,0)$ mm} \\
\cmidrule(lr){2-6}\cmidrule(lr){8-12}
Model & {\scriptsize FWHM$_r$} & {\scriptsize FWHM$_t$} & {\scriptsize FWHM$_z$} & $\Delta\mu_r$ & RMSE & {} & {\scriptsize FWHM$_r$} & {\scriptsize FWHM$_t$} & {\scriptsize FWHM$_z$} & $\Delta\mu_r$ & RMSE \\\midrule
\modelDepth & 1.52 & 1.52 & 1.46 & -0.00 & 1.09 & {} & 1.48 & 1.75 & 1.59 & 0.10 & 1.17 \\
\modelBase & 1.72 & 1.74 & 1.83 & 0.00 & 1.27 & {} & 4.26 & 2.03 & 1.84 & 3.15 & 3.64 \\
\modelSEG{2} & 1.80 & 1.81 & 1.91 & -0.00 & 1.33 & {} & 4.27 & 2.15 & 1.93 & 3.15 & 3.67 \\
\modelSEG{3} & 1.90 & 1.90 & 2.00 & 0.00 & 1.40 & {} & 4.29 & 2.29 & 2.03 & 3.17 & 3.70 \\
\modelSEG{4} & 1.96 & 1.97 & 2.07 & 0.00 & 1.48 & {} & 4.34 & 2.49 & 2.17 & 3.18 & 3.75 \\
\midrule
\multicolumn{12}{@{}l}{\textbf{Intrinsic FWHM: 4 mm}} \\[2pt]
 & \multicolumn{5}{c}{Source at $(0,0,0)$ mm} & {} & \multicolumn{5}{c}{Source at $(100,0,0)$ mm} \\
\cmidrule(lr){2-6}\cmidrule(lr){8-12}
Model & {\scriptsize FWHM$_r$} & {\scriptsize FWHM$_t$} & {\scriptsize FWHM$_z$} & $\Delta\mu_r$ & RMSE & {} & {\scriptsize FWHM$_r$} & {\scriptsize FWHM$_t$} & {\scriptsize FWHM$_z$} & $\Delta\mu_r$ & RMSE \\\midrule
\modelDepth & 1.83 & 1.83 & 1.76 & -0.00 & 1.28 & {} & 1.74 & 2.10 & 1.91 & 0.17 & 1.41 \\
\modelBase & 2.03 & 2.05 & 2.11 & 0.00 & 1.50 & {} & 4.34 & 2.41 & 2.15 & 3.17 & 3.75 \\
\modelSEG{2} & 2.10 & 2.11 & 2.17 & 0.00 & 1.55 & {} & 4.34 & 2.51 & 2.22 & 3.17 & 3.77 \\
\modelSEG{3} & 2.18 & 2.20 & 2.24 & 0.00 & 1.61 & {} & 4.36 & 2.66 & 2.30 & 3.18 & 3.80 \\
\modelSEG{4} & 2.27 & 2.29 & 2.31 & 0.00 & 1.69 & {} & 4.41 & 2.83 & 2.42 & 3.19 & 3.85 \\
\bottomrule\end{tabular}
\end{table}

\subsubsection{Brain-phantom reconstruction}
\label{sec:resultsBrainRecon}

\begin{figure}[htb]
    \centering
    \includegraphics[width=\linewidth]{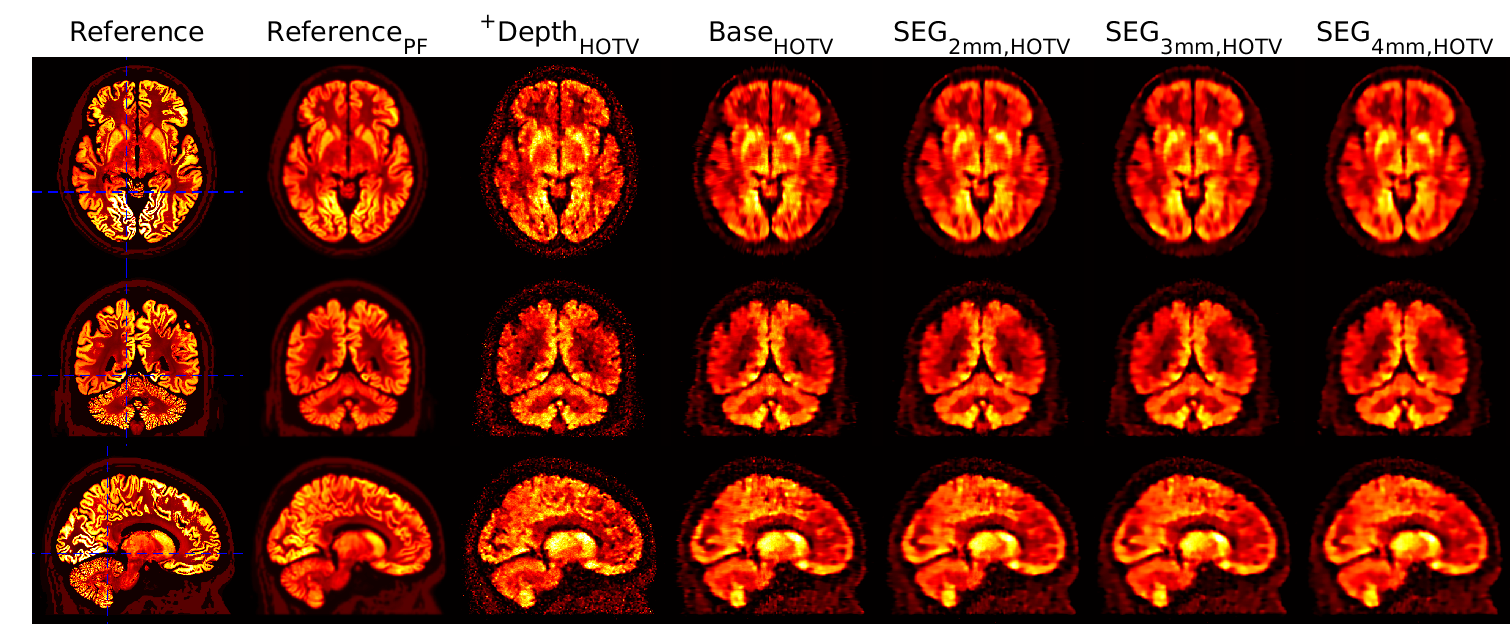}
    \caption{GATE brain T1-MR PET-phantom HOTV reconstructions derived from a high-resolution MR-based activity distribution with 1-mm FWHM intrinsic resolution. }
    \label{fig:brainRecon}
\end{figure}

Brain-phantom HOTV reconstructions at 1-mm intrinsic detector resolution are shown in Fig.~\ref{fig:brainRecon}.  These images were reconstructed with 0.667-mm cubic voxels.  Penalty weights for the DOI groups \modelDepth and the nonDOI others (\modelBase and \modelSEG{p}) were found by golden search giving: DOI group $\lambda_{TV_{1,2}}=0.00052,0.00031$, and nonDOI group $\lambda_{TV_{1,2}}=0.00031,0.00021$, respectively.  In addition, a reference image blurred with a 2.53 mm FWHM Gaussian filter, matching the resolution of the \modelDepth image is shown to facilitate comparison between the models.  Models without measured DOI (\modelBase and the segmented models) exhibited visible blurring and distortion near the brain periphery, whereas \modelDepth reduced these effects.  Quantitative comparisons are summarized in Table~\ref{tab:brain-hotv-metrics}.  Among the evaluated models, \modelDepth yielded the lowest NRMSE and highest PSNR and SSIM.  The resolution-matched reference provides a comparison with the underlying high-resolution activity distribution.

\begin{table}[htb]
    \centering
    \caption{Image quality metrics for the resolution matched reference image and HOTV reconstructions for 1-mm intrinsic resolution data.  Reference\(_\text{PF}\) represents the metrics for reference image with 2.53 mm post-filtering, based on the estimated resolution loss in the \modelDepth model.}
    \label{tab:brain-hotv-metrics}
    \renewcommand{\arraystretch}{0.95}
   \begin{tabular}{@{}lcccc@{}}
        \toprule
        Model & NRMSE & PSNR (dB) & SSIM & Matched FWHM \\
        \midrule
        Reference\(_\text{PF}\) & 0.271 & 24.9 & 0.704 & 2.53 mm\\
        \modelDepth             & 0.389 & 21.7 & 0.521 & 2.55 mm\\
        \modelBase              & 0.422 & 21.0 & 0.450 & 3.39 mm\\
        \modelSEG{2}            & 0.420 & 21.1 & 0.445 & 3.64 mm\\
        \modelSEG{3}            & 0.423 & 21.0 & 0.437 & 3.76 mm\\
        \modelSEG{4}            & 0.428 & 20.9 & 0.427 & 3.95 mm\\
        \bottomrule
    \end{tabular}
\end{table}

\subsection{Projector computational performance}
\label{sec:resultsCompute}

Although the SPLAT forward- and back-projectors are conceptually simple (Fig.~\ref{fig:splat-pseudocode}), the original MATLAB implementation was computationally prohibitive.  We therefore developed optimized C++/CUDA implementations as MATLAB MEX functions, achieving a speedup of more than three orders of magnitude over the original MATLAB prototype in development tests.  Table~\ref{tab:projectorPerformance} summarizes GPU throughput on an NVIDIA RTX 4000 Ada for \(10^7\) list-mode events, four image matrices, and coincidence timing resolutions (CTR) ranging from 600 to 50 ps FWHM.  For a fixed image matrix, throughput generally increased with improved CTR, as the narrower TOF kernels required fewer voxel evaluations.  Throughput decreased with increasing image matrix size.  At 200-ps CTR, the \(125\times125\times250\) (2~mm) grid achieved 5.44 and 6.02 million events/s for forward- and back-projection, respectively, compared with 0.18 and 0.14 million events/s for the \(500\times500\times1000\) (0.5~mm) grid.

\begin{table*}[htb]
\centering
\caption{SPLAT forward/back-projection throughput on an NVIDIA RTX 4000 Ada GPU.  Each test used \(10^7\) list-mode events and assumed 3-mm FWHM spatial detector resolution.  Values are millions of events per second (higher is better).}
\label{tab:projectorPerformance}
\small
\setlength{\tabcolsep}{3pt}
\renewcommand{\arraystretch}{0.95}
\begin{tabular}{@{}l*{7}{C{1.35cm}}@{}}
\toprule
& \multicolumn{7}{c}{CTR (ps FWHM); forward/back-projection throughput}\\
\cmidrule(lr){2-8}
Image matrix
& 600 & 400 & 300 & 200 & 150 & 100 & 50\\
\midrule
$125\times125\times250$  & 2.04/2.30 & 2.81/3.14 & 3.66/4.08 & 5.44/6.02 & 7.22/7.96 & 10.9/11.8 & 86.3/51.0 \\
$250\times250\times500$  & 0.37/0.39 & 0.49/0.51 & 0.64/0.64 & 0.94/0.92 & 1.24/1.19 & 1.87/1.74 & 3.64/3.30 \\
$375\times375\times750$  & 0.14/0.13 & 0.18/0.17 & 0.23/0.21 & 0.34/0.30 & 0.45/0.39 & 0.67/0.57 & 1.29/1.07 \\
$500\times500\times1000$ & 0.07/0.06 & 0.10/0.08 & 0.12/0.10 & 0.18/0.14 & 0.23/0.19 & 0.34/0.27 & 0.68/0.51 \\
\bottomrule
\end{tabular}
\end{table*}

\section{Discussion}
\label{sec:discussion}

The principal contribution of this study is the development of SPLAT, a continuous-domain, event-specific projection model for TOF list-mode PET reconstruction.  In SPLAT, each coincidence is represented by a spatial probability kernel in image space, constructed from the measured event attributes and their associated uncertainties.  This formulation allows the detector measurements to be incorporated directly into consistent forward and adjoint projection operators without requiring detector-coordinate binning.

An important feature of SPLAT is that the event-response kernel can be tailored to the physical characteristics of the detector and the photon interactions being modeled.  For this initial investigation, we chose an anisotropic Gaussian response to describe selected photoelectric coincidence events.  This choice provides a relatively simple representation of detector-localization uncertainties and is consistent with the approximately Gaussian measurement errors prescribed in our detector model.  Propagating these uncertainties through the event geometry produces a three-dimensional response whose center, extent, and orientation depend on the individual coincidence.  More complex physical interactions or detector responses could be accommodated through other probability distributions without changing the underlying continuous-domain formulation.

An important aspect of this study is that the predicted event-response distributions were evaluated directly against the known annihilation positions from GATE Monte Carlo simulations.  This provides a physical reference for evaluating both localization accuracy and the probability assigned to the true annihilation position, independently of image reconstruction or comparisons with other projection models.  For the selected photoelectric coincidences, the Gaussian responses showed little mean localization bias and approximately nominal 95\% and 99\% coverage.  These results support the use of event-specific Gaussian kernels to represent the combined effects of detector measurement uncertainties and photon non-collinearity under the simulated conditions.

The comparisons between projection models demonstrate the importance of incorporating the available detector measurements into the event response.  Measured DOI provided the largest improvement in localization accuracy, primarily by reducing uncertainty in the photon interaction positions and the associated parallax errors.  The advantages of preserving continuous detector coordinates were greatest at the highest intrinsic detector resolutions, where segmentation discards a larger proportion of the available spatial information.  Incorporating energy-dependent correlations and depth-dependent detector resolution provided smaller but consistent additional improvements.  Together, these results demonstrate the value of propagating the detailed detector response through the geometry of each coincidence rather than relying on fixed or spatially averaged resolution models.

The benefits of more accurate event localization were also apparent in the reconstructed images.   In the point-source grid, \modelDepth reduced localization bias and spatial broadening compared with models without measured DOI, particularly toward the radial limits of the field of view.  Similar effects were observed in the brain-phantom reconstructions, where the models without measured DOI exhibited peripheral blurring and distortion.  These observations are consistent with the geometric effects illustrated in Fig.~\ref{fig:splatPSFCompare} and are particularly relevant to small-diameter PET systems, where oblique photon incidence can produce substantial parallax errors.  Some of the systematic displacement observed with the fixed-DOI and segmented models could potentially be reduced through geometric calibration or spatial corrections.  However, such corrections would not recover the event-specific interaction-depth information or eliminate the associated uncertainty.  The reconstruction improvements therefore reflect reductions in both systematic localization errors and spatial broadening.

Improved event localization can increase the information contributed by each accepted coincidence.  The information-equivalent count gain developed here arose naturally from the SPLAT representation of each event as a spatial probability density, whose extent and orientation reflect its localization uncertainty.  For Gaussian responses, the squared spatial concentration provides a simple measure of this uncertainty that can be calculated directly from the event covariance.  Combined with count efficiency, it offers a general approach to estimating the relative information provided by different detector configurations and response models.  Its connection to Fisher information follows from the weak-point-source approximation developed in Section~\ref{sec:projectionValidation}, while the practical calculation also accounts for positron range and finite voxel dimensions.

The proposed measure is also consistent with conventional estimates of TOF sensitivity gain.  For a separable Gaussian response with fixed transverse uncertainty, the squared spatial concentration is inversely proportional to the longitudinal TOF uncertainty.  Relative to a non-TOF response distributed uniformly over an object of diameter \(D\), the resulting information-equivalent count gain is
\begin{equation}
    G_{\mathrm{TOF}}
    \simeq
    \sqrt{\frac{2\ln 2}{\pi}}\,
    \frac{D}{\Delta x}
    \simeq
    0.66\,\frac{D}{\Delta x},
    \qquad
    \Delta x=\frac{c\,\mathrm{CTR}}{2},
\end{equation}
consistent with established TOF variance-gain estimates \cite{tomitani1981image, budinger1983time, schaart2021time}. This relationship provides a useful connection between conventional TOF gain and the information-equivalent count gain introduced here, which extends the comparison to three-dimensional event responses incorporating detector-position, DOI, and correlated measurement uncertainties.  The resulting gains are approximate measures of localization information; their relationship to task-specific detectability and reconstructed-image quality depends on the signal, background distribution, and reconstruction method \cite{barrett1997list, fessler1996spatial, rahmim2013resolution}.

As shown in Section~\ref{sec:projectionValidation}, the proposed measure is motivated by the list-mode Fisher-information framework \cite{barrett1997list, parra1998list}.  For a weak point source in a uniform background, spatially averaged Fisher information is proportional to the accepted count and the mean squared-kernel concentration under the stated approximations, while the practical calculation incorporates Gaussian-equivalent positron-range and voxel-size effects.  For the matched coincidence data used in this study, the count-efficiency ratio is unity, allowing the comparisons to focus on differences in the modeled spatial response.  The resulting gains provide an additional perspective on the benefits of incorporating continuous detector measurements and event-specific uncertainties.  Their relationship to detectability and reconstructed-image quality depends on the signal characteristics, background distribution, and reconstruction method \cite{barrett1997list, fessler1996spatial, rahmim2013resolution}.

Another useful aspect of the formulation is the separate treatment of positron range.  Unlike detector localization uncertainty and photon non-collinearity, positron transport occurs before annihilation and is therefore represented by an independent operator mapping the radioactive-decay distribution to the annihilation distribution.  This distinction allows the event-response model and positron-range model to be developed independently, including for radionuclides with longer positron ranges or spatially varying tissue properties.

Although the Gaussian implementation performed well for the selected photoelectric coincidences, departures from ideal Gaussian behavior were observed in the distribution tails.  In particular, the elevated dispersion of the squared Mahalanobis distance at high intrinsic detector resolution indicates that some event responses are not adequately described by a single Gaussian.  More realistic detector responses may exhibit asymmetry, extended tails, or multiple spatial components associated with detector boundaries and multi-site photon interactions \cite{cloquet2010nongaussian, yao2012lorpdf}.  These effects could be represented within SPLAT using Gaussian mixtures, heavy-tailed distributions such as Student's \(t\), or empirical response kernels derived from detector measurements or Monte Carlo simulations.  Mixture models could also distinguish different photon point-of-first-interaction histories, including photoelectric--photoelectric, Compton--photoelectric, and Compton--Compton coincidences.  Importantly, these extensions would not require changing the underlying continuous-domain integral-operator formulation.

Finally, this investigation used prescribed detector-response characteristics rather than measurements from a physical scanner, allowing the contributions of individual detector measurements to be evaluated under controlled conditions.  The analysis was also restricted to selected photoelectric coincidences in a simplified cylindrical detector geometry.  Further work will be needed to incorporate measured detector responses, more complex photon-interaction histories, and scattered and random coincidences.  The optimized GPU implementation demonstrates that event-specific forward- and back-projection can be performed efficiently, providing a practical basis for these extensions.  The flexibility to incorporate detector-specific response kernels makes SPLAT applicable to a range of PET geometries and detector technologies, particularly systems capable of measuring interaction positions and their uncertainties with high precision.

\section{Conclusions}
\label{sec:conclusions}

We developed SPLAT, a continuous-domain, event-specific projection model for time-of-flight list-mode PET reconstruction.  The model represents each detected coincidence by a spatial response kernel in image space, incorporating measured detector attributes and their uncertainties without requiring discretization of the detector coordinates.  In this study, we implemented SPLAT using anisotropic Gaussian kernels that account for TOF, DOI, photon non-collinearity, and energy-dependent and correlated detector responses.

Monte Carlo validation demonstrated improved event localization compared with matched segmented-detector models.  The largest improvement resulted from incorporating measured DOI, with additional gains from energy-dependent correlations and depth-dependent detector resolution.  Point-source reconstructions confirmed that these improvements can reduce spatial bias and improve localization accuracy, particularly in regions affected by parallax.

The continuous-domain formulation is not restricted to Gaussian responses and provides a flexible framework for incorporating more detailed detector models into PET reconstruction.  These results establish a foundation for applying SPLAT to advanced position-sensitive PET detectors and evaluating the benefits of improved event localization in reconstructed images.  

\section*{Acknowledgments}

Early two-dimensional implementations of the SPLAT projection model and image reconstruction framework were explored by Erik Linghammar and Richard Nunstedt in a master's thesis co-supervised by C.  Ross Schmidtlein and Ida H\"aggstr\"om~\cite{linghammar2025positron}.  This work was supported in part by NIH grant 1R01EB034742-01.  C.  Ross Schmidtlein and Ronglong Fang were supported in part by the MSK Cancer Center Support Grant/Core P30 CA008748.

\bibliographystyle{plainnat}
\bibliography{report}
\end{document}